\documentclass[aps,prb,reprint,superscriptaddress,longbibliography]{revtex4-2}

\usepackage{glossaries}
\usepackage{graphicx}
\usepackage{bm}
\usepackage{hyperref}
\usepackage{color}
\usepackage{comment}
\usepackage{verbatim}
\usepackage{soul}
\usepackage{chemformula}
\usepackage{siunitx}
\usepackage[colorinlistoftodos]{todonotes}

\newacronym{QSL}{QSL}{quantum spin liquid}
\newacronym{SOC}{SOC}{spin orbit coupling}
\newacronym{xanes}{XANES}{x-ray absorption near edge structure}
\newacronym{xmcd}{XMCD}{x-ray magnetic circular dichroism}
\newacronym{rixs}{RIXS}{resonant inelastic x-ray scattering}
\newacronym{br}{BR}{branching ratio}
\newacronym{dft}{DFT}{density functional theory}
\newacronym{aps}{APS}{Advanced Photon Source}
\newacronym{anl}{ANL}{Argonne National Laboratory}
\newacronym{rt}{RT}{room temperature}
\def\cuiro{$\mathrm{Cu_2IrO_3}$}
\def\liiro{$\mathrm{Li_2IrO_3}$}
\def\nairo{$\mathrm{Na_2IrO_3}$}
\def\agliiro{$\mathrm{Ag_3LiIr_2O_6}$}
\def\hliiro{$\mathrm{H_3LiIr_2O_6}$}
\def\ls{$\langle \mathbf{L} \cdot \mathbf{S} \rangle$}
\def\lzsz{$\langle L_z \rangle/\langle S_z \rangle$}

\DeclareSIUnit\planckbar{\text {\ensuremath {\hbar }}}

\DeclareRobustCommand{\element}[1]{\@element#1\@nil}
\def\@element#1#2\@nil{%
  #1%
  \if\relax#2\relax\else\MakeLowercase{#2}\fi}
\makeatother

\begin{document}

\title{Novel electronic state of honeycomb iridate Cu$_2$IrO$_3$ at high pressure}

\author{G. Fabbris}
\email[]{gfabbris@anl.gov}
\affiliation{Advanced Photon Source, Argonne National Laboratory, Lemont, Illinois 60439, USA}

\author{E. H. T. Poldi}
\affiliation{Advanced Photon Source, Argonne National Laboratory, Lemont, Illinois 60439, USA}
\affiliation{Department of Physics, University of Illinois at Chicago, Chicago, Illinois 60607, USA}

\author{S. Sinha}
\affiliation{Department of Physics, University of Florida, Gainesville, Florida 32611, USA}

\author{J. Lim}
\affiliation{Department of Physics, University of Florida, Gainesville, Florida 32611, USA}
\affiliation{Department of Physics, Eastern Illinois University, Charleston, Illinois 61920, USA}

\author{T. Elmslie}
\affiliation{Department of Physics, University of Florida, Gainesville, Florida 32611, USA}

\author{J. H. Kim}
\affiliation{Advanced Photon Source, Argonne National Laboratory, Lemont, Illinois 60439, USA}

\author{A. Said}
\affiliation{Advanced Photon Source, Argonne National Laboratory, Lemont, Illinois 60439, USA}

\author{M. Upton}
\affiliation{Advanced Photon Source, Argonne National Laboratory, Lemont, Illinois 60439, USA}

\author{M. Abramchuk}
\affiliation{Department of Physics, Boston College, Chestnut Hill, Massachusetts 02467, USA}

\author{F. Bahrami}
\affiliation{Department of Physics, Boston College, Chestnut Hill, Massachusetts 02467, USA}

\author{C. Kenney-Benson}
\affiliation{HPCAT, X-ray Science Division, Argonne National Laboratory, Lemont, Illinois 60439, USA}

\author{C. Park}
\affiliation{HPCAT, X-ray Science Division, Argonne National Laboratory, Lemont, Illinois 60439, USA}

\author{G. Shen}
\affiliation{HPCAT, X-ray Science Division, Argonne National Laboratory, Lemont, Illinois 60439, USA}

\author{Y. K. Vohra}
\affiliation{Department of Physics, University of Alabama at Birmingham, Birmingham, Alabama 35294, USA}

\author{R. J. Hemley}
\affiliation{Department of Physics, University of Illinois at Chicago, Chicago, Illinois 60607, USA}
\affiliation{Departments of Chemistry, and Earth and Environmental Sciences, University of Illinois Chicago, Chicago, Illinois 60607, USA}

\author{J. J. Hamlin}
\affiliation{Department of Physics, University of Florida, Gainesville, Florida 32611, USA}

\author{F. Tafti}
\affiliation{Department of Physics, Boston College, Chestnut Hill, Massachusetts 02467, USA}
    
\author{D. Haskel}
\affiliation{Advanced Photon Source, Argonne National Laboratory, Lemont, Illinois 60439, USA}

\date{\today}

\begin{abstract}

\ \ \cuiro\ has attracted recent interest due to its proximity to the Kitaev quantum spin liquid state and the complex structural response observed at high pressures. We use x-ray spectroscopy and scattering as well as electrical transport techniques to unveil the electronic structure of \cuiro\ at ambient and high pressures. Despite featuring a $\mathrm{Ir^{4+}}$ $J_{\rm{eff}}=1/2$ state at ambient pressure, Ir $L_{3}$ edge resonant inelastic x-ray scattering reveals broadened electronic excitations that point to the importance of Ir $5d$-Cu $3d$ interaction. High pressure first drives an Ir-Ir dimer state with collapsed \ls\ and \lzsz, signaling the formation of $5d$ molecular orbitals. A novel $\mathrm{Cu \to Ir}$ charge transfer is observed above \SI{30}{\GPa} at low temperatures, leading to an approximate $\mathrm{Ir^{3+}}$ and $\mathrm{Cu^{1.5+}}$ valence, with persistent insulating electrical transport seemingly driven by charge segregation of $\mathrm{Cu^{1+}}$/$\mathrm{Cu^{2+}}$ ions into distinct sites. Concomitant x-ray spectroscopy and diffraction measurements through different thermodynamic paths demonstrate a strong electron-lattice coupling, with $J_{\rm{eff}}=1/2$ and $\mathrm{Ir^{3+}}$/$\mathrm{Cu^{1.5+}}$ electronic states occurring only in phases 1 and 5, respectively. Remarkably, the charge-transfer state can only be reached if \cuiro\ is pressurized at low temperature, suggesting that phonons play an important role in the inhibiting this phase. These results point to the choice of thermodynamic path across interplanar collapse transition as a key parameter to access novel states in intercalated iridates.

\end{abstract}

\maketitle

\section{Introduction}

The realization that strong spin-orbit coupling can drive electronic correlation and emergent phenomena has revolutionized research on heavy transition metals materials \cite{Witczak-Krempa2014, Rau2016, Cao2018}. Particular attention has been devoted to the emergence of $J_{\rm{eff}} = 1/2$ orbitals in $\mathrm{Ir^{4+}}$ and its consequences \cite{Kim2009, Jackeli2009}, with recent focus centered on the theoretically predicted Kitaev \gls{QSL} state due to its potential use in topologically protected quantum computing \cite{Kitaev2006}. Fundamentally, the Kitaev \gls{QSL} is an analytically solvable Hamiltonian that relies on a combination of a 2D honeycomb structural lattice and bond-directional exchange interaction, with the spin-orbit-driven $J_{\rm{eff}} = 1/2$ state of honeycomb iridates being a strong candidate to realize these conditions \cite{Jackeli2009}. However, stabilizing the Kitaev \gls{QSL} state has proven to be difficult since, in real materials, other magnetic interactions are not fully quenched and compete with the Kitaev exchange \cite{Liu2022}, usually leading to magnetic order as in \liiro\ and \nairo\ \cite{Liu2011, Ye2012, Biffin2014b, Chun2015, Williams2016}. The recent development of intercalated honeycomb iridates, namely \cuiro\ \cite{Abramchuk2017}, \agliiro\ \cite{Bahrami2019}, and \hliiro\ \cite{Kitagawa2018}, has re-ignited the field since magnetic order is suppressed in these materials. However, the intercalation process drives structural disorder and can disrupt the Ir $J_{\rm{eff}}=1/2$ orbital \cite{Kenney2019, Knolle2019, delaTorre2021}, raising questions on whether the Kitaev \gls{QSL} has indeed been stabilized.

\begin{figure}[t]
\includegraphics[width = \linewidth]{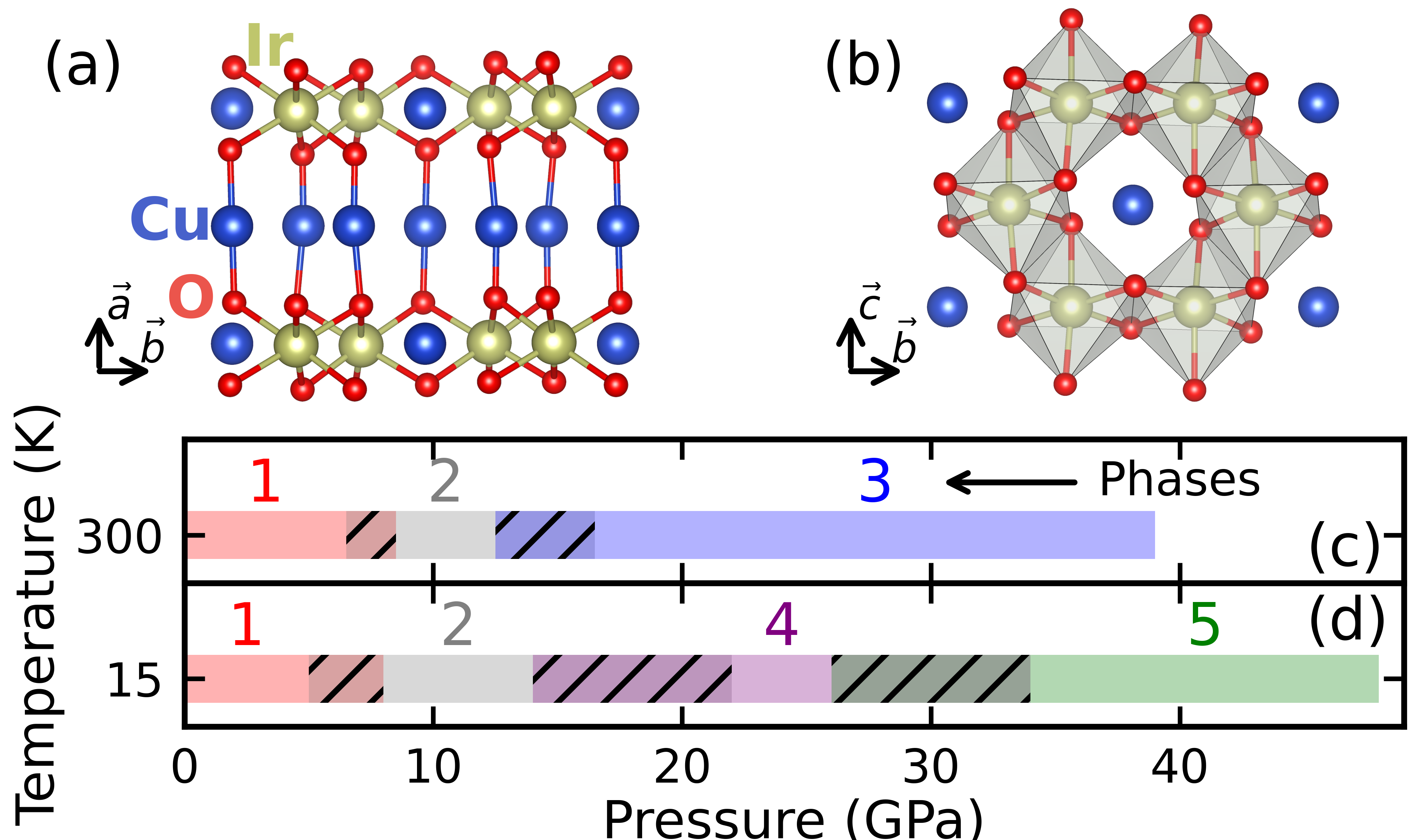}
\caption{(a)\&(b) Crystal structure at ambient pressure featuring Ir honeycomb layers and interplanar O-Cu-O dumbbells. (c)\&(d) High pressure structural phase diagram of \cuiro\ at room and low temperature \cite{Fabbris2021}, respectively, as determined through isotherm measurements. The dashed areas are regions of coexistence of two phases. }
\label{fig1}
\end{figure}

\cuiro\ is distinct to other intercalated iridates by the presence of Cu ions both between and within the Ir honeycomb layers [Figs. \ref{fig1}(a,b)] \cite{Abramchuk2017, Bahrami2022}. This leads to the unusual octahedrally coordinated $\mathrm{Cu^{1+}}$ ions in plane. There is evidence for the presence of magnetic $\mathrm{Cu^{2+}}$ ($\sim 10\%$), which is argued to be nucleated in $\mathrm{Cu^{2+}/Ir^{3+}}$ domains \cite{Kenney2019}. The apparent proximity between these charge states raises the potential for structural control of the electronic and/or magnetic properties. Indeed, \cuiro\ displays a very rich high pressure structural phase diagram \cite{Fabbris2021, Jin2022, Pal2023}, featuring five different crystal structures (phases 1 through 5) and distinct sets of phase transitions at room and low temperatures [Figs. \ref{fig1}(c,d)] \cite{Fabbris2021}. Low pressures (\qty[parse-numbers = false]{\sim 6 - 7.5}{\GPa}) drive Ir-Ir dimerization (phase 2) \cite{Fabbris2021, Jin2022, Pal2023}, similar to reports on \liiro\ polymorphs \cite{Veiga2017, Veiga2019, Hermann2018, Clancy2018} and \agliiro\ \cite{Jin2024}. Higher pressures lead to the collapse of the interplanar distance and a distinct structural temperature dependence, with phase 3 appearing around \SI{15}{\GPa} at \gls{rt}, and phases 4 and 5 around \qtylist{18;30}{\GPa} at low temperature, respectively \cite{Fabbris2021}. While a substantial drop in resistance was observed at onset of phase 3 \cite{Jin2022, Pal2023}, there is little information on if and/or how the electronic structure is coupled to these phase transitions.

In this work, we investigate the relationship between the crystal and electronic structure in \cuiro\ using \gls{xanes}, \gls{xmcd}, \gls{rixs}, x-ray powder diffraction, and electrical transport measurements as a function of pressure and temperature. The data reveal that the Ir-Ir dimerization in phase 2 suppresses the localized $J_{\rm{eff}}=1/2$ state, which is phenomenologically similar to \liiro\ polymorphs \cite{Veiga2017, Veiga2019, Clancy2018, Takayama2019}. Further compression at low temperature drives a $\mathrm{Cu \to Ir}$ electron transfer at the onset of phase 5. Despite the Cu mixed valency, \cuiro\ remains an insulator in phase 5, which appears to be due to charge segregation into $\mathrm{Cu^{1+}/Cu^{2+}}$ sites. No charge transfer occurs at similar pressure at \gls{rt} (phase 3). Remarkably, no electronic or structural phase transition is observed upon cooling from phase 3 (\gls{rt} to \SI{15}{\K}) or warming from phase 5 (\SI{15}{\K} to \gls{rt}). This demonstrates a strong electron-lattice coupling as the charge transfer is intrinsically connected to the phase 5 structure. It also points to an intricate phase stability landscape in \cuiro, with distinct phenomenology driven by the specific thermodynamic path.

\section{Electronic structure of phase 1 at ambient pressure}

We start by addressing the electronic structure of \cuiro\ in phase 1 at ambient pressure. Intercalated honeycomb iridates were conceived to reduce the exchange interaction between the honeycomb layers, while preserving the well defined $J_{\rm{eff}}=1/2$ character seen in \nairo\ and \liiro, leading to a more 2D structural motif that better mimics the Kitaev model \cite{Kitaev2006}. To explore the detailed electronic structure of \cuiro, \gls{xanes} and \gls{xmcd} were measured at the 4-ID-D and \gls{rixs} at the 27-ID beamlines of the \gls{aps}, \gls{anl}. All measurements were performed on powdered samples that were grown as described in ref. \cite{Abramchuk2017}, and \gls{xmcd} was collected using a magnetic field of \SI{4}{T}. Experimental details are described in Appendix \ref{appendix_expdet}. At ambient pressure, \cuiro\ displays the typical $\mathrm{Ir^{4+}}$ \gls{xanes} and \gls{xmcd} spectra at both \SI{300}{K} and \SI{1.5}{K} [Fig. \ref{figdimers}(a-d)], featuring large $L_3$/$L_2$ intensity ratio. Sum rules analysis reveals \ls~=~\qty{2.64 \pm 0.03}{\square\planckbar} and \lzsz~=~$2.4(1)$ \cite{Thole1988, Carra1993, CommentTz}, which is consistent with other honeycomb iridates \cite{Clancy2012, Veiga2017, Clancy2018, delaTorre2021} and, more generally, with $J_{\rm{eff}}=1/2$ Ir oxides \cite{Laguna-Marco2010, Haskel2012, Clancy2012, Laguna-Marco2015}.

\gls{rixs} measurements, however, point to a more complex picture, with meaningful differences in the orbital excitations among the honeycomb iridates. \nairo\ and $\mathrm{\alpha-Li_2IrO_3}$ feature an exciton peak at \qty[parse-numbers = false]{\sim420-450}{\meV}, and a pair of trigonal crystal field split $J_{\rm{eff}} = 3/2$ peaks at about  \qty[parse-numbers = false]{720-830}{\meV} \cite{Gretarsson2013b}. A markedly different excitation spectra is observed in $\mathrm{Ag_3LiIr_2O_6}$, with theoretical modeling pointing to a substantially larger Ir-O hybridization, which is argued to enhance magnetic frustration \cite{delaTorre2021}. In \cuiro, a low energy peak near \SI{0.6}{eV} is clearly observed at the Ir $L_3$, but not at the $L_2$, demonstrating the $J_{\rm{eff}}=1/2$ ground state [Fig. \ref{figdimers}(g)]. However, while the spectra appears to be composed of three peaks, these are largely broadened compared to other honeycomb iridates \cite{Gretarsson2013b, Clancy2018, delaTorre2021, delaTorre2023}. Since measurements were done in powdered \cuiro, this suggests more dispersive $J_{\rm{eff}}$ excitations, which points to an increased delocalization of the $5d$ orbitals. Finally, compared to \cuiro\ and \agliiro, \hliiro\ features electronic excitations that much closer resemble those of $\alpha$-\liiro\ \cite{Gretarsson2013b, delaTorre2021, delaTorre2023}, indicating the $d$-electron-based intercalation ion plays a key role in disrupting the Ir $5d$ orbital.

\section{\element{Ir}-\element{Ir} dimers and the suppression of the \boldmath$J_{\mathrm{eff}} = 1/2$ state in phase 2}

Pressure drives the onset of phase 2 at both room (\qty[parse-numbers = false]{6\pm1.5}{\GPa}) and low (\qty[parse-numbers = false]{7.5\pm1}{\GPa}) temperatures [Figs. \ref{fig1}(c,d)], featuring Ir-Ir structural dimers \cite{Fabbris2021}. The emergence of phase 2 is closely correlated with a disruption of the $J_{\rm{eff}}=1/2$ state, as seen by the reduction of \ls\ and \lzsz\ [Fig. \ref{figdimers}(e,f)], as well as the suppression of the low energy excitation ($J_{\rm{eff}}=3/2$, $<$~\SI{1}{eV}) [Fig. \ref{figdimers}(g)]. These results are similar to those observed at the dimerization of $\alpha \mathrm{-}$ and $\beta \mathrm{-Li_2IrO_3}$ \cite{Veiga2017, Clancy2018, Hermann2018, Takayama2019, Veiga2019, vanVeenendaal2022}. However, contrary to \liiro\ \cite{Clancy2018, Takayama2019}, new low energy \gls{rixs} excitations are not seen in \cuiro, instead an increase in the incoherent continuum is observed [Fig. \ref{figdimers}(g)]. Quasimolecular orbitals are likely also stabilized in \cuiro, but the influence of hybridization with Cu $3d$ orbitals masks their low energy excitations. Notably, however, while the exact nature of the quasimolecular orbital cannot be determined, there is strong evidence for the persistent importance of \gls{SOC}. Both \ls\ and \lzsz\ remain sizeable. But more strikingly, the \gls{rixs} excitation near \SI{2.2}{eV}, which at ambient pressure is related to the \gls{SOC} split $t_{2g}$ band \cite{Gretarsson2013b}, not only persists at the Ir $L_3$ edge, but remains absent at the Ir $L_2$ [Fig. \ref{figdimers}(g)]. These results are consistent with a picture that includes both large hopping between the dimerized Ir and large \gls{SOC}, as described for other iridates \cite{Mazin2012, Revelli2019, Wang2019b, vanVeenendaal2022}. Finally, an increase in resistance is observed in phase 2 at \gls{rt} [Fig. \ref{figphase345}(j)], consistent with a previous report \cite{Jin2022}, and in agreement with a larger band gap obtained from \gls{dft} calculations \cite{Fabbris2021}.

\section{Electronic structure of phase 3}

\begin{figure}[t]
\includegraphics[width = \linewidth]{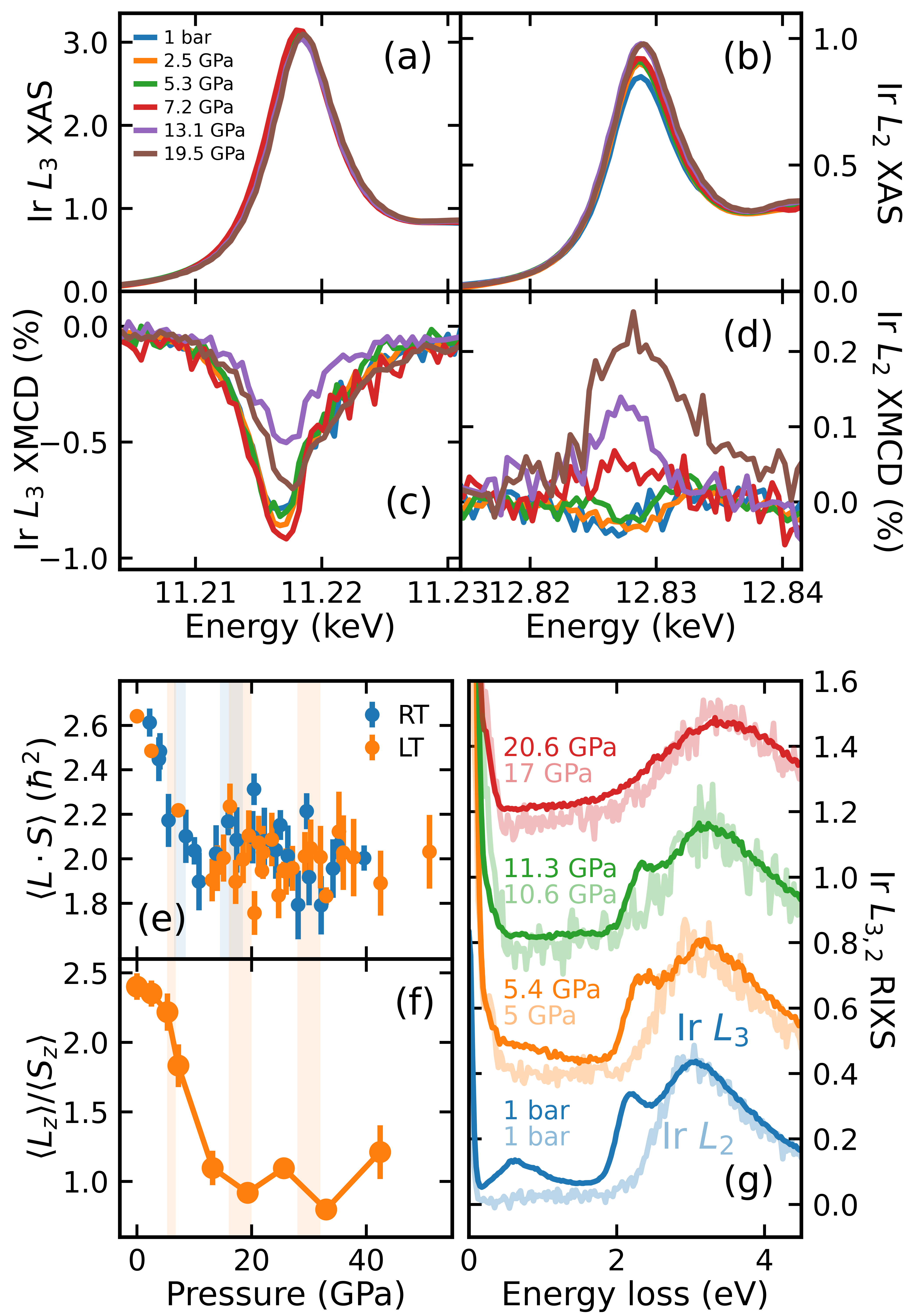}
\caption{Electronic properties of \cuiro\ across the dimerization transition. (a)-(d) Pressure dependence of the Ir $L_{3,2}$ XANES and XMCD spectra at low temperature (\SI{1.5}{K} $<$ T $<$ \SI{15}{K}). Sum rules analysis of the XANES and XMCD yields the (e) average spin-orbit coupling and (f) orbital to spin moment ratio. (g) RIXS pressure dependence at the Ir $L_3$ (\SI{11.215}{keV} incident energy) and $L_2$ (\SI{12.821}{keV}) at room temperature. Vertical shaded areas in panels (e)\&(f) correspond to the structural phase transition regions at \gls{rt} (blue) and low temperature (orange) \cite{Fabbris2021}.}
\label{figdimers}
\end{figure}

\begin{figure*}[t]
\includegraphics[width = \linewidth]{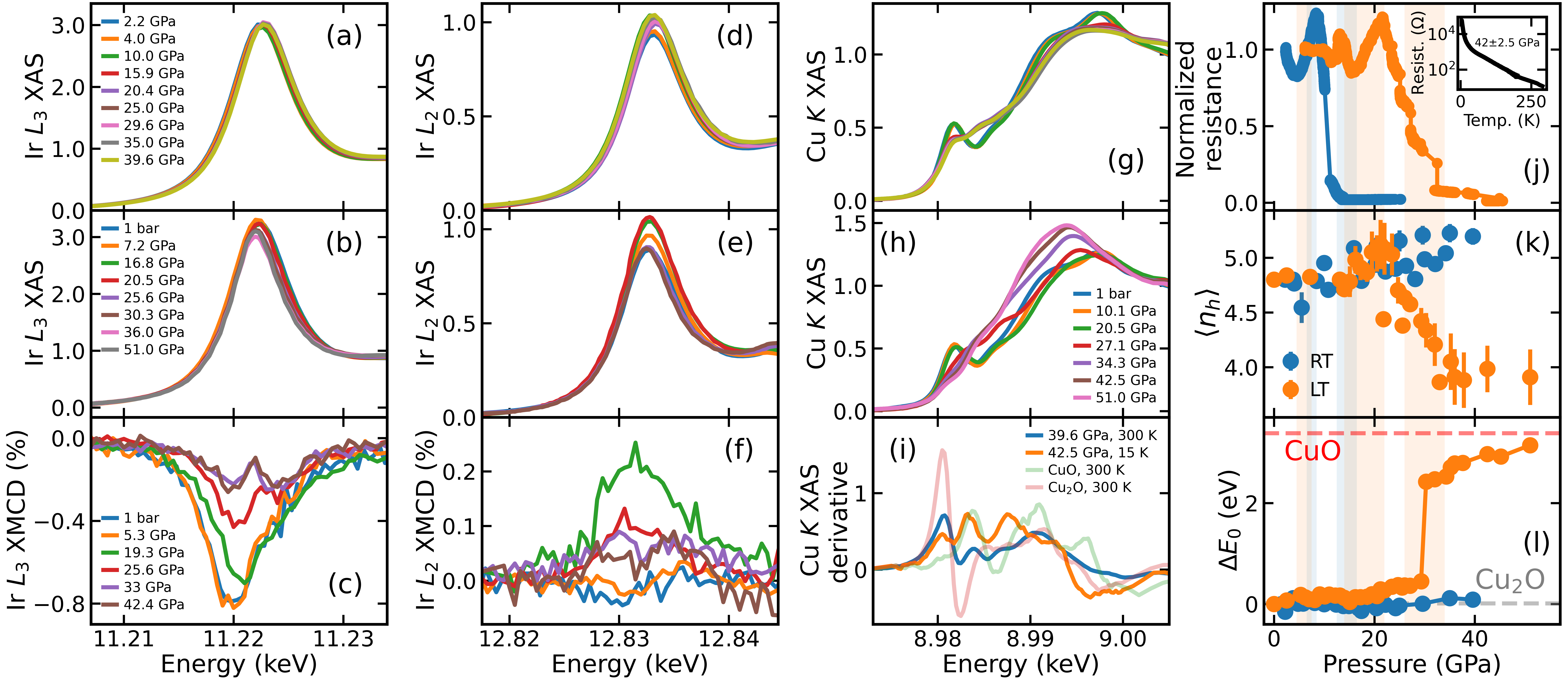}
\caption{Electronic properties of \cuiro\ across phases 3, 4 and 5. Panels (a-c) and (d-f) display the pressure dependence of the room and low temperature \gls{xanes} as well as low temperature \gls{xmcd} taken at the Ir $L_3$ and $L_2$ edges, respectively. (g)\&(h) Cu $K$ edge \gls{xanes} at room and low temperature, respectively. Low temperature was kept between \SI{1.5}{K} and \SI{15}{K}. Panels (d)\&(g) follow the same legend as (a). (i) Comparison between the Cu $K$ edge \gls{xanes} derivative of Cu references (\ch{Cu2^{1+}O}) and \ch{Cu^{2+}O}) and \cuiro\ at room and low temperature taken at approximately the same pressure (corresponding to phases 3 and 5). (j) Resistance pressure dependence at room and low temperatures. The resistance was normalized to its value at ambient pressure (\qtylist{42;584}{\ohm} at room and low temperatures, respectively). A substantial drop in resistance is seen near the onset pressure of phases 3 and 5. Inset: resistance temperature dependence upon warming at \qty[parse-numbers = false]{42\pm2.5}{\GPa} showing that phase 5 remains insulating. (k)\&(l) Number of $5d$ holes ($\langle n_h \rangle$) and Cu $K$ edge shift at room and low temperatures extracted from \gls{xanes} data. An electron transfer from Cu to Ir occurs only at the onset of phase 5. Vertical shaded areas in panels (j-l) correspond to the structural phase transition regions at \gls{rt} (blue) and low temperature (orange) \cite{Fabbris2021}.}
\label{figphase345}
\end{figure*}

While the onset of phase 2 primarily affects the Ir $5d$ state, the Cu orbitals become more active at higher pressures. Curiously, the large reduction in interplanar distance ($d$) at the onset of phase 3 ($\Delta d/d \approx 8 \%$, Fig. \ref{pxrd} and ref. \cite{Fabbris2021}) is not reflected in the Ir $L_{3,2}$ \gls{xanes}, with no distinct changes to \ls\ or $5d$ occupation [Figs. \ref{figdimers}(e) and \ref{figphase345}(k)]. However, the \gls{rixs} excitation near \SI{2.2}{\eV} is largely suppressed in phase 3 [Fig. 2(g)], becoming indistinguishable from the intense continuum of excitations that develops in this phase and extends to zero energy loss. A similar increase in the incoherent background is seen at the Ir $L_2$ \gls{rixs} [Fig. \ref{figdimers}(g)]. These results suggest that the electronic structure of phase 3 features strongly hybridized Ir $5d$-Cu $3d$ orbitals that destroy the quasimolecular orbitals of phase 2, and is consistent with the substantial drop in resistance [Fig. \ref{figphase345}(j)]. The electrical resistance temperature dependence points to a possible metallicity near the onset of phase 3, with increasing semiconducting behavior at higher pressures (see Appendix \ref{appendix_resistance}) \cite{Jin2022}, but, as shown in section \ref{tdep}, phase 3 is sensitive to the thermodynamic path, complicating the interpretation of these results. The nature of these orbitals is potentially interesting as \ls\ remains sizeable, suggesting that, despite the collapse of the $J_{\rm{eff}}$ state and quasimolecular orbitals, \gls{SOC} still plays an important role. No energy shift is observed in the Cu $K$ edge spectra, indicating that its modification is likely a structural response due to strongly modified interplanar O-Cu-O dumbbells, likely without an increase in the Cu coordination.

\section{Phase 5 novel $\mathbf{Ir^{3+}}$-$\mathbf{Cu^{1.5+}}$ electronic state}

At low temperature, the onset of phase 5 is marked by a reduction in the Ir $L_{3,2}$ white line area and shift in the Cu $K$ edge energy (Fig. \ref{figphase345}). Assuming that 80\% of Ir are 4+ at low pressures ($\langle n_h \rangle = 4.8$, where $\langle n_h \rangle$ is the number of $5d$ holes) \cite{Kenney2019}, the reduction in the white line area implies $\langle n_h \rangle \sim 3.9$ above \SI{30}{\GPa} [Fig. \ref{figphase345}(k)], i.e. an approximate $\mathrm{Ir^{3+}}$ valence state. The Cu valence can be extracted by the position of the absorption edge, which is defined as the maximum of the \gls{xanes} first derivative. While the multi-featured Cu $K$ edge of \cuiro\ largely complicates a quantitative analysis, a clear change in the first derivative is seen at the onset of phase 5 [Fig. \ref{figphase345}(i)], leading to a discontinuous jump in position of the absorption edge [Fig. \ref{figphase345}(l)]. Notably, the Cu $K$ edge taken in phase 3 is very distinct, and features a lower energy edge [Fig. \ref{figphase345}(i)]. The sharpness of this jump is artificial as the maximum moves from the first to the second peak in the \gls{xanes} derivative [Fig. \ref{figphase345}(i)]. Nevertheless, the increase in energy implies that Cu lost electrons, thus demonstrating a novel Cu $\to$ Ir electron transfer in \cuiro\ that occurs only in phase 5. The substantial difference in the electronic structure of phases 3 and 5 indicates that these phases likely feature distinct Cu environments, despite similar collapsed interplanar distances and stability pressure range.

The $\mathrm{Ir^{3+}}$ state implies that the nominal Cu valence is 1.5+. Metallic electronic transport would be expected if this partial valence was only due to the delocalization of the Cu $3d$ bands. However, while a large drop in resistance is observed at pressures near the phase 5 onset, its temperature dependence points to a persistent electronic gap [inset of Fig. \ref{figphase345}(j)]. Although the maximum of the Cu $K$ edge derivative shifts in phase 5, a sizeable spectral weight is observed at the $\mathrm{Cu^{1+}}$ energy [Fig. \ref{figphase345}(i)], indicating that the average $\mathrm{Cu^{1.5+}}$ is formed by a mixture of 2+ and 1+ ions. The $\mathrm{Ir^{3+}}$ $5d^6$ orbital features fully occupied $t_{2g}$ states, with a large octahedral crystal field [\qty[parse-numbers = false]{\sim3.5}{\eV}, Fig. \ref{figdimers}(g)]. Therefore, while increased Cu/Ir hybridization leads to a lower resistance in phase 5, the $\mathrm{Ir^{3+}}$ configuration combined with the Cu charge segregation results in the insulating behavior. Interestingly, this scenario is consistent with \gls{dft} calculations, despite phase 5 being different from the predicted crystal structure \cite{Fabbris2021}. Finally, the localized $\mathrm{Cu^{2+}}$ ions in phase 5 could drive magnetism. In fact, the small antiferromagnetic response of \cuiro\ at ambient pressure is attributed to the $\mathrm{Cu^{2+}/Ir^{3+}}$ minority domains \cite{Abramchuk2017, Kenney2019}. However, no \gls{xmcd} signal was observed at the Cu $K$ edge (Appendix \ref{appendix_cuxmcd}). While this suggests that Cu is not ferromagnetically ordered, it does not discard a possible antiferromagnetic state.

\begin{figure*}
\includegraphics[width = \linewidth]{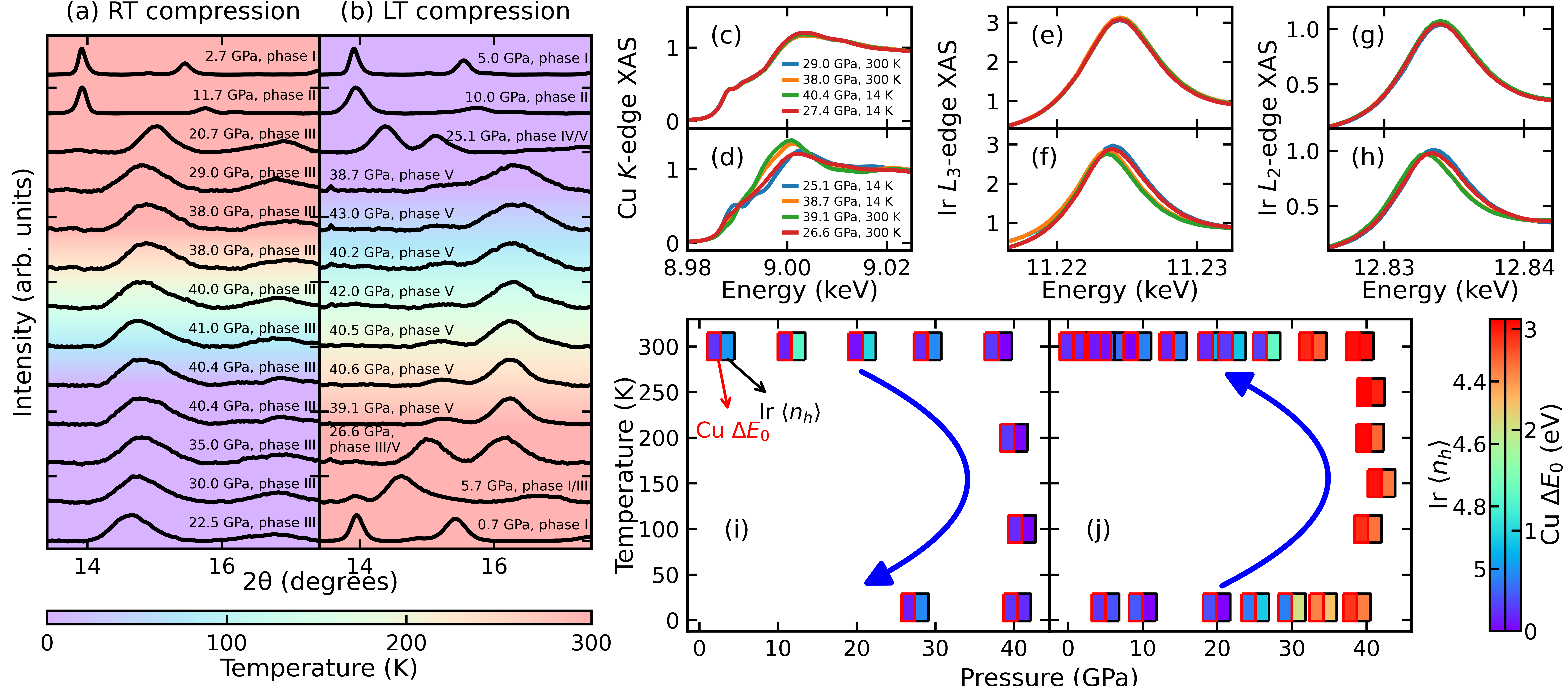}
\caption{Structural and electronic properties of \cuiro\ through two distinct thermodynamical paths. (a), (c), (e) and (g) display x-ray diffraction as well as Cu $K$ and Ir $L_{3,2}$ edge \gls{xanes} measured while pressurizing at \gls{rt} and depressurizing at \SI{15}{\K}, respectively, while (b), (d), (f), and (h) display the same measurements taken via the reverse path. (i)\&(j) Number of $5d$ holes ($\langle n_h \rangle$) and the shift in Cu $K$ edge energy are shown for each path.}
\label{figtdep}
\end{figure*}

Copper delafossites feature a similar crystal structure pressure dependence to \cuiro, albeit with a triangular lattice \cite{Mackenzie2017}. Interestingly, while a pressure-driven interplanar collapse appears to be a general behavior in these compounds \cite{Xu2010, Xu2016, Garg2018, Levy2020, Lawler2021}, only \ch{CuFeO2} is known to feature a similar valence transition \cite{Xu2010, Xu2016}. There is no established explanation for this behavior, but it suggests that the valence transition depends on the relative chemical potential of the related ions. This raises questions on the high pressure behavior of other intercalated iridates. Recently, an unidentified structural transition was reported in \agliiro\ around \qty[parse-numbers = false]{12 - 16.1}{\GPa} at \gls{rt}, which occurs concomitantly with a change in sign of the magnetoresistance \cite{Jin2024}. This phenomenology is similar to that seen in \cuiro\ \cite{Jin2022}, and points to a potential valence instability.

\section{Thermodynamic path dependence of the high pressure electronic structure}
\label{tdep}

The dramatic temperature dependence of the high pressure electronic structure of \cuiro\ led us to explore its thermodynamics through concomitant powder x-ray diffraction and Ir $L_{3,2}$ as well as Cu $K$ edge \gls{xanes} collected at the HPCAT 16-BM-D beamline at the \gls{aps}, \gls{anl} (see Appendix \ref{appendix_expdet} for experimental details). Two temperature/pressure paths were taken: 1) pressure was applied at \SI{300}{\K} up to \SI{38}{\GPa}, followed by a nearly isobaric cool down to \SI{14}{\K} (pressure was maintained between \qtylist[list-units = single]{38;41}{\GPa}), and low temperature decompression; 2) sample was cooled down to \SI{14}{\K} keeping the pressure below \SI{5}{\GPa}, then pressurized to \SI{38.7}{\GPa} at low temperature, followed by a nearly isobaric warm up to \SI{300}{\K} (pressure was kept between \qtylist[list-units = single]{38.7;43}{\GPa}), and \SI{300}{\K} decompression. These different paths drive distinct sets of phase transitions. The compression in path 1) reproduces the phase transitions seen in our previous work \cite{Fabbris2021}, stabilizing phase 3 beyond \qtylist[parse-numbers=false]{\sim 18}{\GPa} with collapsed interplanar distance [Fig. \ref{figtdep}(a)]. High pressure cool down preserves phase 3, which remains stable upon low temperature pressure release to at least \SI{22.5}{\GPa}. Path 2) again leads to the same set of reported low-temperature phases [Fig. \ref{figtdep}(b)] \cite{Fabbris2021}. Remarkably, phase 5 is also preserved upon warming up to \SI{300}{\K}. Pressure release at  \SI{300}{\K} leads to a phase 5$\to$3 transition at \qtylist[parse-numbers=false]{27 \pm 3}{\GPa}, but no signs of phase 4. The dimerized phase 2 is also not detected on pressure release. The remarkable phase diagram of \cuiro\ is closely followed by the electronic structure [Fig. \ref{figtdep}(c-j)]. The Cu$\to$Ir electron transfer occurs only within phase 5, with the reversal to the approximate Ir$^{4+}$/Cu$^{1+}$ configuration at the phase 5$\to$3 transition, and the $J_{\rm{eff}}=1/2$ state features only in phase 1, again including on pressure release through path 2). These results highlight a strong electron-lattice coupling in \cuiro.

The \gls{rt} phase diagram of \cuiro\ is not settled in the literature \cite{Fabbris2021, Jin2022, Pal2023}(there is only one report on the low temperature structures \cite{Fabbris2021}). While there is an overall agreement on the presence of a dimerized Ir-Ir and interplanar collapsed structures (phases 2 and 3), different critical and coexistence pressures have been reported. A key difference amongst these studies is the use of different pressure media, which likely leads to different hydrostaticity. The absence of phase 2 upon decompression from phase 5 further points to the importance of the non-hydrostatic shear in the phase stability of \cuiro, since decompression leads to large pressure gradients. The \cuiro\ growth method may also be relevant to the high pressure phase stability, as recent developments have led to less structural disorder \cite{Haraguchi2024}. Additionally, our results suggest that previous transport measurements did not probe the properties of phases 4 or 5, as pressure was applied at \gls{rt} \cite{Jin2022, Pal2023}. Although our data [Fig. \ref{figphase345}(j)] and theoretical calculations indicate that \cuiro\ is an insulator in phase 5, due to the distinct pressure media used in these measurements, concomitant electrical transport and x-ray diffraction measurements are needed to verify this result.

The processes that stabilize phases 3, 4, 5 remain an important open topic given their distinct electronic properties. The absence of a phase 3$\to$5 (5$\to$3) transition upon cooling (warming) at high pressure prevents determining whether one of these is the ground state. The onset of phase 4 at a similar pressure to phase 3 indicates that its presence is important in stabilizing phase 5. Both of these phases feature a collapsed interplanar distance, which is likely associated with modifications in the O-Cu-O dumbbells (note that this distance is nearly pressure-independent in phases 1 and 2) \cite{Fabbris2021}. Such interplanar dumbbells typically feature strong low energy phonons \cite{Abramchuk2018, Lawler2021}. Therefore, it is likely that the stability of these phases is closely tied to the vibrational amplitude of the dumbbell phonons. This scenario implies that a similar phenomenology will occur in \hliiro\ and \agliiro. Alternatively, low temperature might stifle the nucleation and growth of phase 3, stabilizing instead phase 4. It would be interesting to probe whether the speed of compression affects this transition \cite{Fisch2015} (in this work, we typically pressurized at a rate of \qtylist[parse-numbers=false, per-mode=symbol]{\sim 50}{\GPa\per\hour}, stopping at each pressure point for about \qtylist[parse-numbers=false]{10-30}{\minute}). Further exploration of the \cuiro\ thermodynamic landscape is needed, not only to clarify the role and stability of phase 4, but also to search for novel phenomena.

\section{Conclusion}

This work explored the pressure dependence of the electronic properties of \cuiro. At ambient pressure, an enhanced delocalization of $5d$ $J_{\rm{eff}}=1/2$ orbitals is observed, likely driven by hybridization with the in-plane Cu $3d$ orbitals. High pressure data unveils a remarkable electronic phase diagram. The dimerized Ir-Ir bonds of phase 2 lead to a collapse of the $J_{\rm{eff}}=1/2$ state, likely stabilizing molecular orbitals with persistent $J_{\rm{eff}}$ character. Higher pressures lead to the collapse of the interplanar distances, and to large changes in the electronic structure. While the \gls{rt} phase 3 feature dominant $\mathrm{Ir^{4+}}$/$\mathrm{Cu^{1+}}$ ions, the low temperature phase 5 stabilizes $\mathrm{Ir^{3+}}$/$\mathrm{Cu^{1.5+}}$. The insulating behavior of phase 5 appears to originate in a charge segregation of $\mathrm{Cu^{1+}}$ and $\mathrm{Cu^{2+}}$, raising the prospect of antiferromagnetic order. A remarkable temperature dependence is seen at high pressures with phase 5 only being stabilized upon cold compression, but remaining stable upon warming to \SI{300}{\K}. The richness of the \cuiro\ phase diagram motivates further work. Additionally, it highlights the importance of investigating other intercalated iridates, and the role of the thermodynamic path in driving novel phenomena.

\begin{acknowledgments}
We thank Richard Ferry for the support during experiments at HPCAT. This research used resources of the \gls{aps}, a U.S. Department of Energy (DOE) Office of Science user facility at \gls{anl} and is based on research supported by the U.S. DOE Office of Science-Basic Energy Sciences, under Contract No. DE-AC02-06CH11357. Portions of this work were performed at HPCAT (Sector 16), \gls{aps}, \gls{anl}. HPCAT operations are supported by DOE-NNSA's Office of Experimental Sciences. Helium and neon pressure media were loaded at GeoSoilEnviroCARS (The University of Chicago, Sector 13), \gls{aps}, \gls{anl}. GeoSoilEnviroCARS was supported by the National Science Foundation – Earth Sciences (EAR – 1634415). The work at Boston College was funded by the U.S. Department of Energy, Office of Basic Energy Sciences, Division of Physical Behavior of Materials under award number DE-SC0023124. Work at the at the University of Florida was supported by National Science Foundation (NSF) DMREF-2118718. The University of Illinois Chicago effort was supported by NSF grants DMR-2119308 and DMR-2118020, and by the DOE-NNSA cooperative agreement DE-NA-0003975 (Chicago/DOE Alliance Center, CDAC).
\end{acknowledgments}

\appendix

\section{Experimental details}
\label{appendix_expdet}

\subsection{X-ray spectroscopy} \label{xasdetails}

\Gls{xanes} and \gls{xmcd} experiments were performed at the 4-ID-D beamline of the \gls{aps}, \gls{anl}. Data were collected in transmission geometry at the Ir $L_{2,3}$ and Cu $K$ edges as a function of pressure at room and low (\SI{1.5}{\K} $\leq$ T $\leq$ \SI{15}{\K}) temperatures. A \SI{4}{T} magnetic field was applied to the sample during \gls{xmcd} scans. Magnetic field and temperature were controlled using a \SI{6.5}{T} LHe cooled magnet system. For ambient pressure measurements, \cuiro\ powder was brushed onto Kapton tapes, which were then stacked to generate the desired absorption edge jump ($\sim 0.7$). High pressure was generated using a CuBe cell fitted with a set of partially perforated plus mini anvils of \SI{300}{\um} culet diameter \cite{Haskel2008}. Pressure was controlled in-situ using He gas membranes. Ruby spheres were used as manometer \cite{Dewaele2008}. Helium was used as pressure media and loaded using the GSECARS gas loading facility \cite{Rivers2008}. Monochromatic x-rays were generated using a Si(111) double crystal monochromator. The beam was focused to \qtyproduct[product-units = power]{100 x 250}{\um} using a toroidal mirror, and further reduced to \qtyproduct[product-units = power]{50 x 50}{\um} using slits. X-ray harmonics were rejected by both detunning the monochromator second crystal, and a Pd-coated flat mirror at 3.1 mrad. Circular x-ray polarization for \gls{xmcd} measurements was obtained using a \SI{500}{\um} diamond phase plate. Data analysis was performed using the Larch package \cite{Newville2013}.

\begin{figure}[b]
    \centering
    \includegraphics[width=\linewidth]{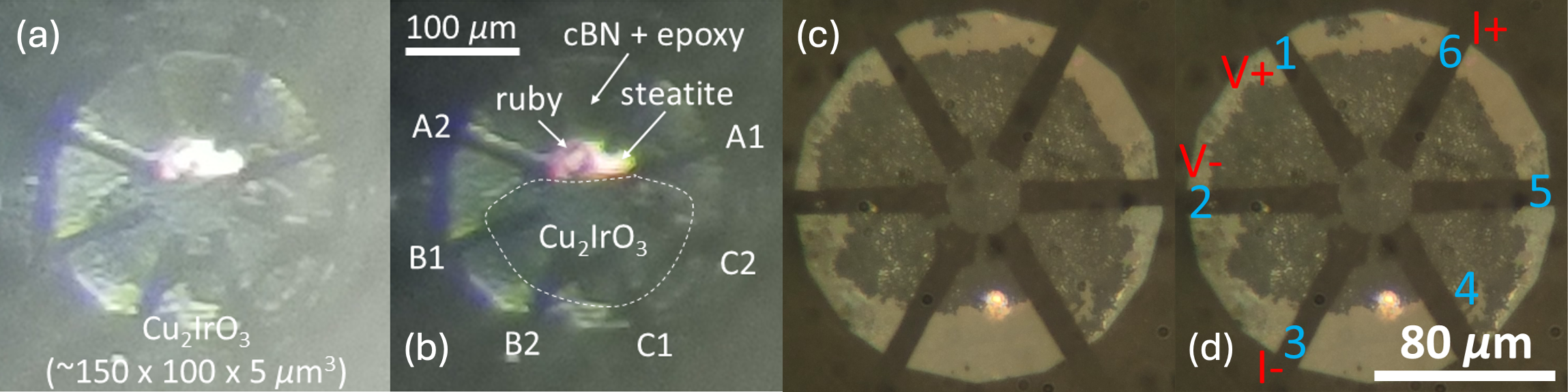}
    \caption{Pictures of the sample configuration for \cuiro\ transport measurements. (a,b) and (c,d) are setup for room and low temperature compression measurements, respectively. (b)\&(d) are the same as (a)\&(c) but include details explaining the electrode configuration.}
    \label{fig:ChiDAC}
\end{figure}

\subsection{Resonant inelastic x-ray scattering}

\begin{figure}[b]
\includegraphics[width = \linewidth] {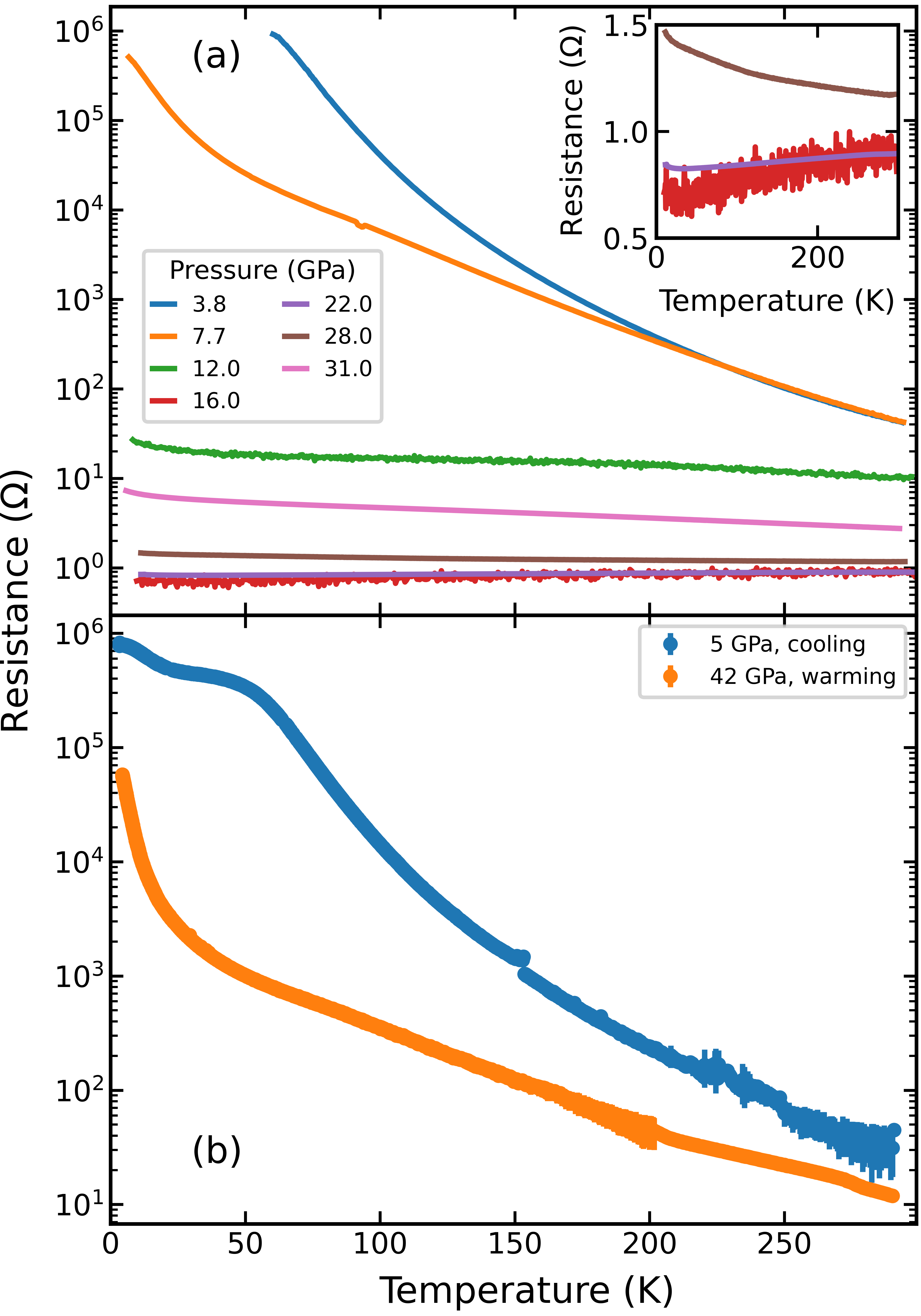}
\caption{\cuiro\ electrical transport temperature dependence at high pressure. (a) Pressure was always applied at room temperature, and electrical transport was measured on warming. (b) Data were collected on cooling at \qty[parse-numbers = false]{5\pm2}{\GPa} (blue points), followed by compression at 15 K. Resistance was then measured at \qty[parse-numbers = false]{42\pm3}{\GPa} on warming (orange points).}
\label{resistance}
\end{figure}

\Gls{rixs} was measured at the Ir $L_{2,3}$ as a function of pressure at room temperature. The experiments were performed at the MERIX instrument located at the 27-ID beamline of the \gls{aps}, \gls{anl}. Incoming x-rays were monochromatized using a combination of high-heat-load diamond (111) and four bounce Si(400) monochromators, and focused to about \qtyproduct[product-units = power]{40 x 25}{\um} (full width half maximum) in the horizontal and vertical directions at the sample position. Inelastic scattering was measured using a Rowland geometry with 2~m diameter, and employing Si(4 4 8) and Ge(0 6 10) spherical analyzers for the Ir $L_{3}$ and $L_2$ edges, respectively. This setup yielded a total \gls{rixs} resolution (full width half maximum) of \SI{90.4}{\meV} at the Ir $L_3$, and \SI{84.3}{\meV} at the $L_2$. Ambient pressure measurements were performed using a \cuiro\ pellet. High pressure was generated using a Princeton-type symmetric diamond anvil cell fitted with regular anvils of \SI{300}{\um} culet diameter. Spurious signal from the diamond anvils can substantially affect the \gls{rixs} spectrum \cite{Kim2016b}. This contamination was reduced by using a gasket-in/gasket-out geometry in which both incoming and outgoing x-rays reach the sample through a Be gasket. Ruby fluorescence was used to calibrate pressure \cite{Dewaele2008}, and neon served as pressure media.

\subsection{Temperature dependent concomitant x-ray spectroscopy and powder diffraction}

Concomitant x-ray powder diffraction as well as Ir $L_{2,3}$ and Cu $K$ edge \gls{xanes} were collected at the HPCAT 16-BM-D beamline of the \gls{aps}, \gls{anl}. High pressure was generated using Princeton-type symmetric cells fitted with a set of partially perforated and regular diamond anvils of \SI{300}{\um} culet diameter. Both the Au lattice constant and ruby fluorescence were employed as pressure calibrants \cite{Holzapfel2001, Ragan1992, Dewaele2008}. Diffraction was collected using x-ray wavelength $\lambda$ = \SI{0.6888}{\AA} (E = \SI{18}{\keV}) in order to facilitate the \gls{xanes} measurements taken at the Ir $L_{2,3}$ (\SI{12.284}{\keV} and \SI{11.214}{\keV}) and Cu $K$ (\SI{8.979}{\keV}) edges.  Measurements were performed using a LHe flow cryostat, with pressure applied in-situ through He gas membranes. Helium was used as pressure medium, and was loaded using the GSECARS gas loading facility \cite{Rivers2008}. The 2D images from the MAR3450 detector were converted to 1D diffractograms using the Dioptas software \cite{Prescher2015}, which was also used to mask diamond Bragg peaks, as well as correct for the diamond and seat absorption. \gls{xanes} was measured in transmission geometry using ion chambers to detect the incident and transmitted x-ray intensity.

\begin{figure*}
\includegraphics[width = 15 cm] {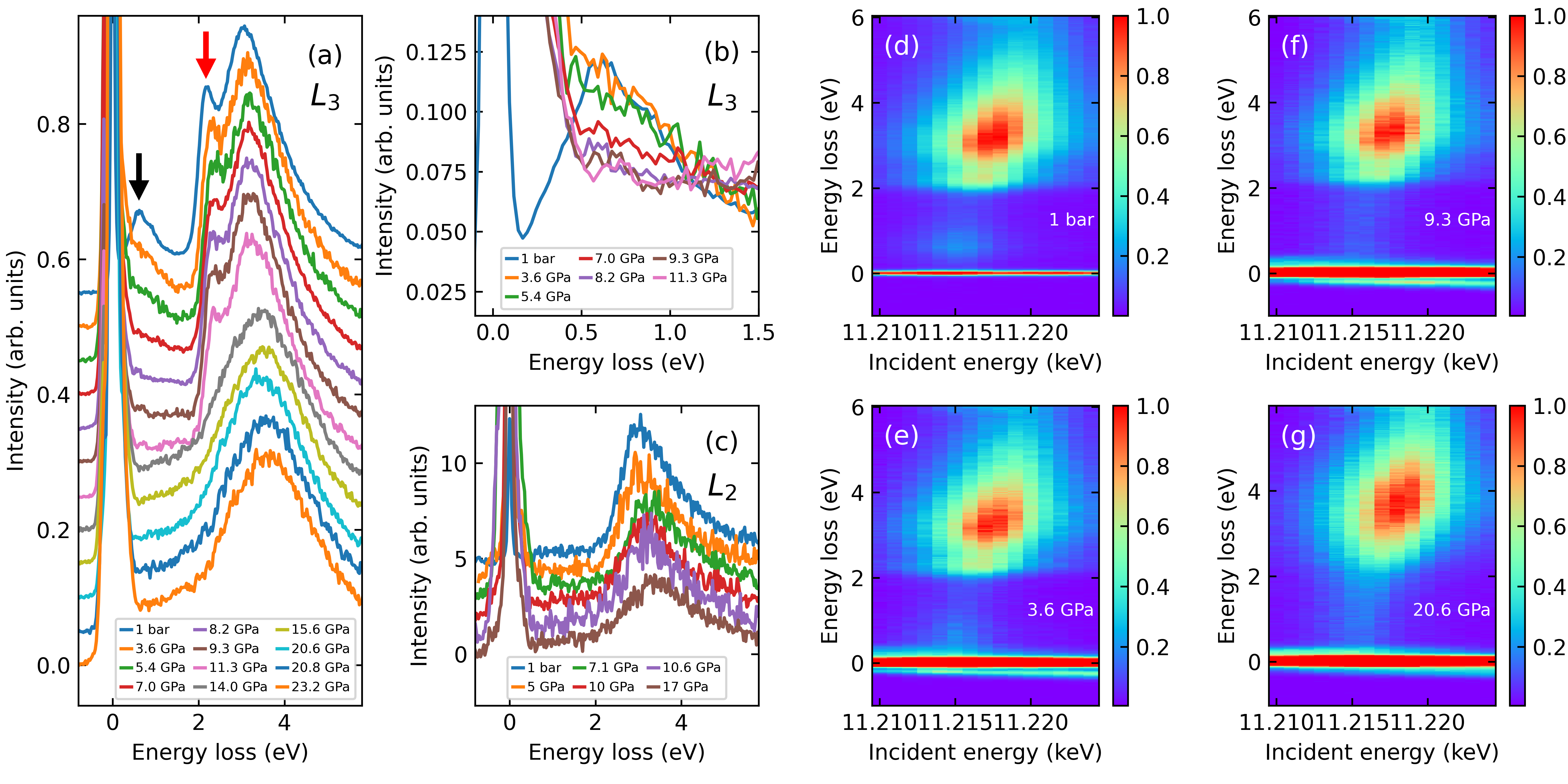}
\caption{\cuiro\ Ir $L_{3,2}$ \gls{rixs} pressure dependence at room temperature. (a) Inelastic signal collected using \SI{11.215}{\keV} incident x-ray energy. Note that the ambient pressure data were collected outside the pressure cell, hence the narrower elastic line. (b) The $J_{\mathrm{eff}} = 1/2 \to 3/2$ excitation is strongly suppressed upon the onset of the dimerized phase 2. (c) RIXS signal from the Ir $L_2$ (\SI{12.821}{\keV}). (d-g) Incident energy dependence of the Ir $L_3$ at selected pressures. While the $J_{\mathrm{eff}} = 1/2 \to 3/2$ excitation is suppressed by pressure, a clear spectral weight remains around \SI{11.215}{\keV} up to at least \SI{20.6}{\GPa} .}
\label{rixs}
\end{figure*}

\subsection{Electrical transport}

\subsubsection{Room-temperature compression}

A powdered \ch{Cu2IrO3} chunk was kept inside an argon-filled glove box with \ch{O2} levels below \SI{0.5}{ppm}. A sample with dimensions of approximately \qtyproduct[product-units = power]{150 x 50 x 5}{\um}, as shown in Fig.~\ref{fig:ChiDAC}, was placed in a gas membrane-based diamond anvil cell (ChicagoDAC from Almax-EasyLab) within one or two hours outside of the glove box, along with a ruby (\SI{20}{\um} in diameter) for pressure calibration \cite{chijioke_ruby_2005}. A designer-diamond anvil with a \SI{20}{\um} central flat, featuring eight symmetrically embedded tungsten electrodes, was used along with a \SI{500}{\um} culet diameter anvil on the opposite side~\cite{Weir_DesignerAnvil_2000}. A 316 stainless steel metal gasket was pre-indented from approximately \SI{150}{\um} to \SI{25}{\um} in thickness, with a hole ($\sim$\SI{140}{\um} in diameter) filled with a cBN-epoxy mixture and soapstone (steatite), which acted as both an electrical insulator and a pressure-transmitting medium. After loading the sample, the cell was placed in a Physical Property Measurement System (PPMS) by Quantum Design. The electrode configuration used was B2: +V, C2: -V, A1: +I, B1: -I [Fig.~\ref{fig:ChiDAC}(b)]. A current of \SI{10}{\uA} was used in all the measurements shown. Resistance was measured in a four-probe arrangement using a Keithley 6221 DC current source and a Keithley 2182a nanovoltmeter configured for ``delta mode." The instruments were set to voltage mode rather than resistance mode. The compliance voltage (maximum voltage allowed to read) was set to \SI{10}{\V}. The temperature was changed at a rate of approximately \qty[per-mode = symbol]{1.5}{\K\per\minute} for each temperature sweep between pressure changes at room temperature.

\subsubsection{Low-temperature compression}

A gas-membrane-driven diamond anvil cell (OmniDAC from Almax-EasyLab) with two opposing diamond anvils (\SI{160}{\um} and \SI{500}{\um} central flats) was used. One of these anvils was a designer-diamond anvil (\SI{160}{\um} central flat) with six symmetrically deposited tungsten microprobes encapsulated in high-quality-homoepitaxial diamond~\cite{Weir_DesignerAnvil_2000}. Stainless steel was used as gasket, as pre-indented to a thickness of around \qtyrange[range-units = single]{20}{25}{\um}. A cBN-epoxy mixture along with steatite was used for insulation and as the pressure medium. A micro-sized \ch{Cu2IrO3} powder sample, with dimensions of approximately \qtyproduct[product-units = power]{140 x 100 x 5}{\um}, was then loaded into the diamond anvil cell within one or two hours outside of the glove box, along with a ruby (\SI{15}{\um} in diameter) for pressure calibration~\cite{chijioke_ruby_2005}. The cell screws were then used to secure the sample and apply an initial pressure of \SI{3}{\GPa}. After attaching the membrane to the cell, a small amount of helium gas pressure (\SI{0.3}{\bar}) was added to the membrane, and the screws were removed.

The gas-membrane-driven diamond anvil cell was then placed inside a customized continuous-flow cryostat (Oxford Instruments). A home-built optical system, attached to the bottom of the cryostat, was used for the visual observation of the sample and measurement of the ruby fluorescence. Initially, the sample was cooled down to \SI{3}{K} at \SI{3}{\GPa}. Pressure was then applied at approximately \SI{15}{\K}, reaching \SI{45}{\GPa}, and the sample was subsequently cooled down to \SI{4}{\K} and warmed up to room temperature while maintaining the pressure between \qty[parse-numbers = false]{40-45}{\GPa}. To reach the mega-ohm range at low temperatures, we used a \SI{1}{\uA} current source. A Keithley delta mode was employed in the same manner as in the room temperature compression experiment. We used a 100 V range in the voltage mode, which showed noise in the low resistance range below \SI{1000}{\ohm}. Upon warming from \SI{200}{K}, we switched to a \SI{10}{\V} range, resulting in reduced noise. Further details of the nonhydrostatic high-pressure resistivity technique, including a photograph using a designer-diamond anvil, are provided in Ref.~\cite{Lim2021}.

\begin{figure*}
\includegraphics[width = 16 cm] {figS4.png}
\caption{\cuiro\ x-ray scattering and spectroscopy data taken through two thermodynamic routes. Room temperature compression (a), followed by nearly isobaric cooldown at \qty[parse-numbers = false]{39\pm1.5}{\GPa}(b), then low temperature (\SI{14}{\K}) pressure release (c). Low temperature (\SI{14}{\K}) compression (d), followed by nearly isobaric warmup around \qty[parse-numbers = false]{40\pm2}{\GPa} (e), then room temperature pressure release (f). (g)\&(h) Pressure dependence of the interplanar distance extracted from the (002) reflection through the room and low temperature compression paths, respectively. (i)\&(j) Ir $5d$ \ls\ measured through the same thermodynamic paths as in (g)\&(h), respectively. The combined data taken through isothermal measurements at room and low temperature are shown in  gray [same data as in Fig. 2(e)].}
\label{pxrd}
\end{figure*}

\section{Additional Electrical Transport Data}
\label{appendix_resistance}

Figure \ref{resistance} displays the temperature dependent resistance measurements that were performed through both room and low (\SI{15}{\K}) compression. The room temperature compression data [Fig. \ref{resistance}(a)] was binned to \SI{0.5}{\K} steps to improve noise level. The low temperature compression measurements employed a low precision temperature sensor, which resulted in digitized temperature steps with multiple resistance measurements per temperature. Fig. \ref{resistance}(b) shows the average of these measurements with error bar being the standard deviation.

The onset of phase 3 (\qty[parse-numbers = false]{14.5\pm2}{\GPa}) at room temperature leads to a sharp drop in resistance [Fig. \ref{figphase345}(k)]. Remarkably, near that transition the temperature dependence suggests a weakly metallic behavior [inset of Fig. \ref{resistance}(a)], but further compression leads to an increase in resistance and re-entry of an insulating/semiconducting state. This behavior is consistent with previous work \cite{Jin2022}. However, given the complex dependence on the thermodynamic path presented in this work, it is unclear whether \cuiro\ retains phase 3 upon cool down near \SI{15}{GPa}, and therefore whether it is truly metallic.

\begin{figure*}
\includegraphics[width = 16 cm] {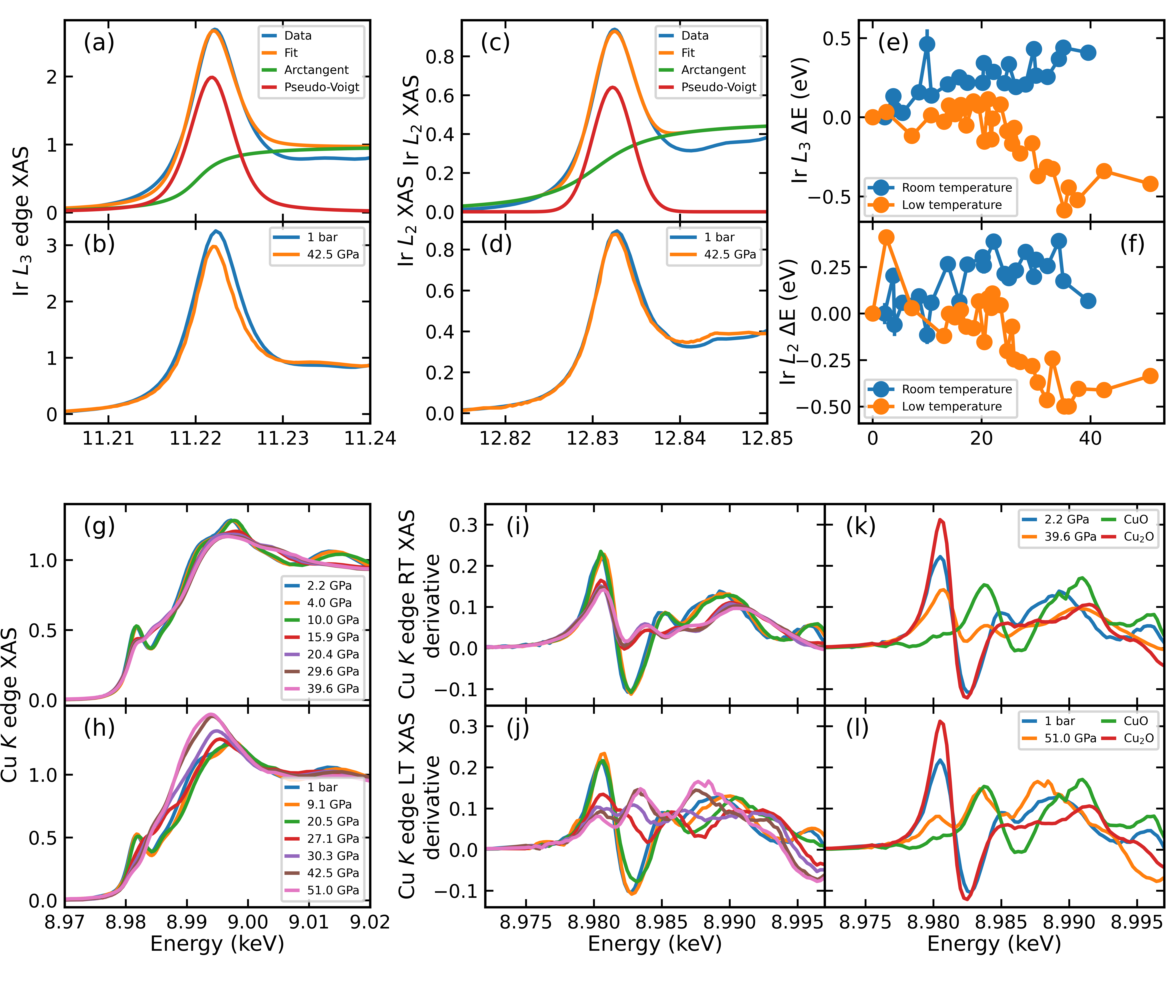}
\caption{Analysis of the \cuiro\ \gls{xanes} data. (a-d) Example of the Ir $L_{3,2}$ fitting and pressure dependence. (e)\&(f) Position of the white line of the Ir $L_3$ and $L_2$ edges extracted from fitting, respectively. (g)\&(h) Cu $K$ edge \gls{xanes} taken during room and low temperature compression, respectively.  (i)\&(j) First derivative of the data in (g)\&(h), the absorption edge is defined as the energy of the maximum of the first derivative. (k)\&(l) Comparison between the lowest and highest pressures of the data shown in (i)\&(j) with data from $\mathrm{Cu^{2+}O}$ and $\mathrm{Cu^{1+}_2O}$ standards.}
\label{xanes}
\end{figure*}

\section{Additional Resonant Inelastic X-ray Scattering Data}
\label{appendix_rixs}

The full Ir $L_{3,2}$ \gls{rixs} dataset is presented in Fig. \ref{rixs}. The characteristic $J_{\mathrm{eff}}=1/2 \to 3/2$ excitation is clearly seen at the $L_3$ edge at ambient pressure, and, as expected, absent at the $L_2$. Scattering from the high pressure environment increases the width of the elastic peak, whose tail overlaps with the $J_{\mathrm{eff}}$ excitation. Nevertheless, a clear suppression of the excitation is seen across the dimerization transition, similar to data from \ch{Li2IrO3} \cite{Takayama2019, Clancy2018}. Interestingly, the peak near \SI{2.2}{eV} [red arrow in Fig. \ref{rixs}(a)] only completely vanished around the onset of phase 3 (\qty[parse-numbers = false]{14.5\pm2}{\GPa}) \cite{Fabbris2021}, and leads to a substantial incoherent excitation continuum that is clearly seen in Fig. \ref{rixs}(g). The $L_2$ edge data also show an increased continuum in phase 3 [Fig. \ref{rixs}(c)]. These results suggest that spin-orbit coupling remains relevant in the dimerized phase (albeit without a $J_{\mathrm{eff}}=1/2$ state), but it is quenched in phase 3, likely due to increased hybridization.

\section{Interplanar distance dependence on the thermodynamic path and recovery of \boldmath$J_{\mathrm{eff}} = 1/2$ state on pressure release}
\label{appendix_distance}

Further details of the structural transitions seen in \cuiro\ for different thermodynamic paths are presented in Fig. \ref{pxrd}. Compression at both room  and low temperature lead to the previously reported phase transitions \cite{Fabbris2021}. This is further evidenced by the interplanar distance pressure dependence [Fig. \ref{pxrd} (g,h)]. The onsets of both phases 3 and 5 are marked by an interplanar distance collapse, with the low temperature phase 5 displaying a shorter distance than the room temperature phase 3 \cite{Fabbris2021}. The phase 5 to 3 transition observed on the cold-compression/warm-release pathway clearly demonstrate that phase 5 indeed has a smaller interplanar distance than phase 3, being thus correlated with their distinct Ir/Cu valence states. The strong electron-lattice coupling is also evident by the Ir \ls\ behavior [Fig. \ref{pxrd}(i,j)], which includes a recovery of the $J_{\mathrm{eff}}=1/2$ state on pressure release.

\section{X-ray absorption spectroscopy data analysis}
\label{appendix_xas}

In order to employ the \gls{xanes} and \gls{xmcd} sum rules analysis at the Ir $L$-edges, the area of the \gls{xanes} white line (peak near the absorption edge) must be extracted \cite{Thole1988, Carra1993}. To this end, we fit the Ir $L_{3,2}$ \gls{xanes} with a combination of arctangent step function and a pseudo-Voigt peak, an example of this fit is displayed in Figs. \ref{xanes}(a,b). In order to estimate systematic errors, we average the area of the pseudo-Voigt with an integral of the data minus the step function, and use their difference as errors.

The Ir valence transition in phase 5 is demonstrated by the $\langle n_h \rangle$ extracted through the Ir $L_{3,2}$ \gls{xanes} [Fig. \ref{figphase345}(k)]. Such transition is also expected to shift the white line position to lower energies \cite{Clancy2012, Laguna-Marco2015}. Figures \ref{xanes}(e,f) display the position of the Ir $L_{3,2}$ white line as a function of pressure. A clear shift is observed in both edges at the onset of phase 5, providing further evidence for the $\mathrm{Ir^{3+}}$ state.

The Cu $K$ edge position, defined as the absorption inflection point, is a signature of the Cu valence state, with the \ch{Cu$^{1+}_2$O} and \ch{Cu$^{2+}$O} edges at \SI{8980.4}{eV} and \SI{8983.7}{eV}, respectively [Fig. \ref{xanes}(k,l)]. While large changes in the \gls{xanes} are seen at the onset of phase 3, the energy of the edge moves only slightly to higher energies (\qty[parse-numbers=false]{\lesssim 0.1}{\eV}), remaining much closer to the $\mathrm{Cu^{1+}}$ reference [Figs. \ref{xanes}(g,k)]. In contrast, phase 5 leads to a clear shift in spectral weight to similar energies as seen in the $\mathrm{Cu^{2+}}$ reference [Figs. \ref{xanes}(h,l)]. But, notably, the Ir 4+ $\to$ 3+ transition [Fig. \ref{figphase345}(k)] implies that Cu is nominally 1.5+ due to the 2/1 stochiometry. It is unclear whether this Cu mixed valence state corresponds to distinct $\mathrm{Cu^{1+}}$/$\mathrm{Cu^{2+}}$ sites (as in a charge order insulator), or a semi-filled band (band insulator). However, the persistence of a clear peak near the 1+ energy suggests that the former scenario is stabilized [Fig. \ref{xanes}(j)].

\section{\element{Cu} $K$ edge x-ray magnetic circular dichroism}
\label{appendix_cuxmcd}

\begin{figure}
\includegraphics[width = \linewidth] {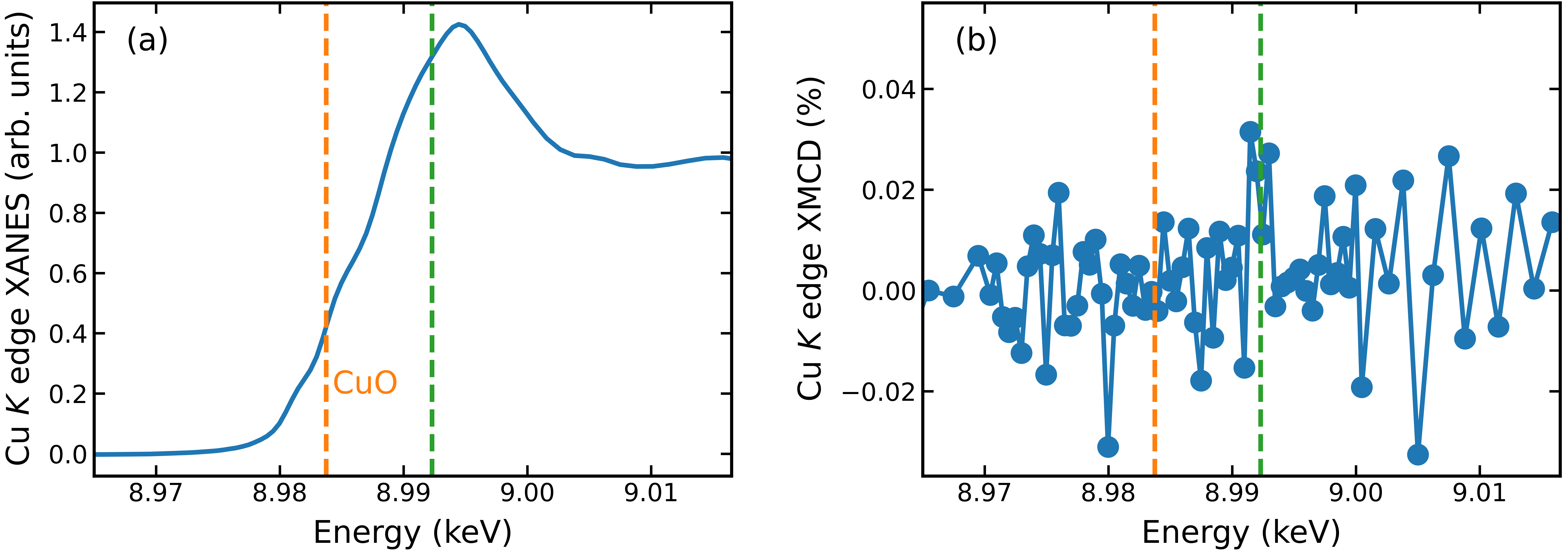}
\caption{\cuiro\ Cu $K$ edge \gls{xanes} (a) and \gls{xmcd} (b) at \SI{42.5}{\GPa} and \SI{1.6}{\K}  (phase 5). The orange dashed line corresponds to the absorption edge position for $\mathrm{Cu^{2+}}$.}
\label{cuxmcd}
\end{figure}

As discussed in the manuscript and Appendix \ref{appendix_xas}, there is evidence for the charge segregation of $\mathrm{Cu^{1+}}$ and $\mathrm{Cu^{2+}}$ ions in phase 5. The $\mathrm{Cu^{2+}}$ $3d^9$ orbitals have a local moment and could lead to magnetic order. In fact, the magnetic signal observed in \cuiro\ at low temperatures is likely due to the $\mathrm{Cu^{2+}/Ir^{3+}}$ minority domains \cite{Abramchuk2017, Kenney2019}. The $K$ edge \gls{xmcd} signal is typically small ($\sim 0.1-0.5 \%$ of the absorption jump) since it relies on the small orbital moment of the $p$ valence states \cite{vanderLaan2014}. There is no clear expectation for the size of the \gls{xmcd} signal for $\mathrm{Cu^{2+}}$, since it typically orders antiferromagnetically, but ferromagnetic Cu metal normally yields an \gls{xmcd} amplitude of $\sim 0.2 \%$ \cite{Nagamatsu2004}. Figure \ref{cuxmcd} shows Cu $K$ edge \gls{xmcd} data taken in phase 5 (\SI{42.5}{\GPa} and \SI{1.6}{\K}) using the same setup described in Appendix \ref{xasdetails}.  No clear \gls{xmcd} signal is observed at the $\sim 0.1 \%$ level, suggesting that the \cuiro\ phase 5 is not ferromagnetic. A small peak is observed at roughly the same energy as the white line (green dashed line in Fig. \ref{cuxmcd}), which is consistent with the paramagnetic signal reported in $\mathrm{Bi_2Sr_2CaCu_2O_{8+y}}$ \cite{Ivanov2018}. However, given its small amplitude and narrow width ($\lesssim 2$ eV), it is unclear if it is a real signal.

\bibliography{refs}

\begin{thebibliography}{64}%
\makeatletter
\providecommand \@ifxundefined [1]{%
 \@ifx{#1\undefined}
}%
\providecommand \@ifnum [1]{%
 \ifnum #1\expandafter \@firstoftwo
 \else \expandafter \@secondoftwo
 \fi
}%
\providecommand \@ifx [1]{%
 \ifx #1\expandafter \@firstoftwo
 \else \expandafter \@secondoftwo
 \fi
}%
\providecommand \natexlab [1]{#1}%
\providecommand \enquote  [1]{``#1''}%
\providecommand \bibnamefont  [1]{#1}%
\providecommand \bibfnamefont [1]{#1}%
\providecommand \citenamefont [1]{#1}%
\providecommand \href@noop [0]{\@secondoftwo}%
\providecommand \href [0]{\begingroup \@sanitize@url \@href}%
\providecommand \@href[1]{\@@startlink{#1}\@@href}%
\providecommand \@@href[1]{\endgroup#1\@@endlink}%
\providecommand \@sanitize@url [0]{\catcode `\\12\catcode `\$12\catcode
  `\&12\catcode `\#12\catcode `\^12\catcode `\_12\catcode `\%12\relax}%
\providecommand \@@startlink[1]{}%
\providecommand \@@endlink[0]{}%
\providecommand \url  [0]{\begingroup\@sanitize@url \@url }%
\providecommand \@url [1]{\endgroup\@href {#1}{\urlprefix }}%
\providecommand \urlprefix  [0]{URL }%
\providecommand \Eprint [0]{\href }%
\providecommand \doibase [0]{https://doi.org/}%
\providecommand \selectlanguage [0]{\@gobble}%
\providecommand \bibinfo  [0]{\@secondoftwo}%
\providecommand \bibfield  [0]{\@secondoftwo}%
\providecommand \translation [1]{[#1]}%
\providecommand \BibitemOpen [0]{}%
\providecommand \bibitemStop [0]{}%
\providecommand \bibitemNoStop [0]{.\EOS\space}%
\providecommand \EOS [0]{\spacefactor3000\relax}%
\providecommand \BibitemShut  [1]{\csname bibitem#1\endcsname}%
\let\auto@bib@innerbib\@empty
\bibitem [{\citenamefont {Witczak-Krempa}\ \emph {et~al.}(2014)\citenamefont
  {Witczak-Krempa}, \citenamefont {Gang}, \citenamefont {Baek},\ and\
  \citenamefont {Leon}}]{Witczak-Krempa2014}%
  \BibitemOpen
  \bibfield  {author} {\bibinfo {author} {\bibfnamefont {W.}~\bibnamefont
  {Witczak-Krempa}}, \bibinfo {author} {\bibfnamefont {C.}~\bibnamefont
  {Gang}}, \bibinfo {author} {\bibfnamefont {K.~Y.}\ \bibnamefont {Baek}},\
  and\ \bibinfo {author} {\bibfnamefont {B.}~\bibnamefont {Leon}},\ }\bibfield
  {title} {\bibinfo {title} {Correlated quantum phenomena in the strong
  spin-orbit regime},\ }\href
  {https://doi.org/10.1146/annurev-conmatphys-020911-125138} {\bibfield
  {journal} {\bibinfo  {journal} {Annual Review of Condensed Matter Physics}\
  }\textbf {\bibinfo {volume} {5}},\ \bibinfo {pages} {57} (\bibinfo {year}
  {2014})}\BibitemShut {NoStop}%
\bibitem [{\citenamefont {Rau}\ \emph {et~al.}(2016)\citenamefont {Rau},
  \citenamefont {Lee},\ and\ \citenamefont {Kee}}]{Rau2016}%
  \BibitemOpen
  \bibfield  {author} {\bibinfo {author} {\bibfnamefont {J.~G.}\ \bibnamefont
  {Rau}}, \bibinfo {author} {\bibfnamefont {E.~K.-H.}\ \bibnamefont {Lee}},\
  and\ \bibinfo {author} {\bibfnamefont {H.-Y.}\ \bibnamefont {Kee}},\
  }\bibfield  {title} {\bibinfo {title} {Spin-orbit physics giving rise to
  novel phases in correlated systems: Iridates and related materials},\ }\href
  {https://doi.org/10.1146/annurev-conmatphys-031115-011319} {\bibfield
  {journal} {\bibinfo  {journal} {Annual Review of Condensed Matter Physics}\
  }\textbf {\bibinfo {volume} {7}},\ \bibinfo {pages} {195} (\bibinfo {year}
  {2016})}\BibitemShut {NoStop}%
\bibitem [{\citenamefont {Cao}\ and\ \citenamefont
  {Schlottmann}(2018)}]{Cao2018}%
  \BibitemOpen
  \bibfield  {author} {\bibinfo {author} {\bibfnamefont {G.}~\bibnamefont
  {Cao}}\ and\ \bibinfo {author} {\bibfnamefont {P.}~\bibnamefont
  {Schlottmann}},\ }\bibfield  {title} {\bibinfo {title} {The challenge of
  spin-orbit-tuned ground states in iridates: a key issues review},\ }\href
  {https://doi.org/10.1088/1361-6633/aaa979} {\bibfield  {journal} {\bibinfo
  {journal} {Reports on Progress in Physics}\ }\textbf {\bibinfo {volume}
  {81}},\ \bibinfo {pages} {042502} (\bibinfo {year} {2018})}\BibitemShut
  {NoStop}%
\bibitem [{\citenamefont {Kim}\ \emph {et~al.}(2009)\citenamefont {Kim},
  \citenamefont {Ohsumi}, \citenamefont {Komesu}, \citenamefont {Sakai},
  \citenamefont {Morita}, \citenamefont {Takagi},\ and\ \citenamefont
  {Arima}}]{Kim2009}%
  \BibitemOpen
  \bibfield  {author} {\bibinfo {author} {\bibfnamefont {B.~J.}\ \bibnamefont
  {Kim}}, \bibinfo {author} {\bibfnamefont {H.}~\bibnamefont {Ohsumi}},
  \bibinfo {author} {\bibfnamefont {T.}~\bibnamefont {Komesu}}, \bibinfo
  {author} {\bibfnamefont {S.}~\bibnamefont {Sakai}}, \bibinfo {author}
  {\bibfnamefont {T.}~\bibnamefont {Morita}}, \bibinfo {author} {\bibfnamefont
  {H.}~\bibnamefont {Takagi}},\ and\ \bibinfo {author} {\bibfnamefont
  {T.}~\bibnamefont {Arima}},\ }\bibfield  {title} {\bibinfo {title}
  {{Phase-Sensitive Observation of a Spin-Orbital Mott State in
  $\mathrm{Sr_2IrO_4}$}},\ }\href {https://doi.org/10.1126/science.1167106}
  {\bibfield  {journal} {\bibinfo  {journal} {Science}\ }\textbf {\bibinfo
  {volume} {323}},\ \bibinfo {pages} {1329} (\bibinfo {year}
  {2009})}\BibitemShut {NoStop}%
\bibitem [{\citenamefont {Jackeli}\ and\ \citenamefont
  {Khaliullin}(2009)}]{Jackeli2009}%
  \BibitemOpen
  \bibfield  {author} {\bibinfo {author} {\bibfnamefont {G.}~\bibnamefont
  {Jackeli}}\ and\ \bibinfo {author} {\bibfnamefont {G.}~\bibnamefont
  {Khaliullin}},\ }\bibfield  {title} {\bibinfo {title} {{Mott Insulators in
  the Strong Spin-Orbit Coupling Limit: From Heisenberg to a Quantum Compass
  and \uppercase{K}itaev Models}},\ }\href
  {https://doi.org/10.1103/PhysRevLett.102.017205} {\bibfield  {journal}
  {\bibinfo  {journal} {Physical Review Letters}\ }\textbf {\bibinfo {volume}
  {102}},\ \bibinfo {pages} {017205} (\bibinfo {year} {2009})}\BibitemShut
  {NoStop}%
\bibitem [{\citenamefont {\uppercase{K}itaev}(2006)}]{Kitaev2006}%
  \BibitemOpen
  \bibfield  {author} {\bibinfo {author} {\bibfnamefont {A.}~\bibnamefont
  {\uppercase{K}itaev}},\ }\bibfield  {title} {\bibinfo {title} {{Anyons in an
  exactly solved model and beyond}},\ }\href
  {https://doi.org/10.1016/j.aop.2005.10.005} {\bibfield  {journal} {\bibinfo
  {journal} {Annals of Physics}\ }\textbf {\bibinfo {volume} {321}},\ \bibinfo
  {pages} {2} (\bibinfo {year} {2006})}\BibitemShut {NoStop}%
\bibitem [{\citenamefont {Liu}\ \emph {et~al.}(2022)\citenamefont {Liu},
  \citenamefont {Chaloupka},\ and\ \citenamefont {Khaliullin}}]{Liu2022}%
  \BibitemOpen
  \bibfield  {author} {\bibinfo {author} {\bibfnamefont {H.}~\bibnamefont
  {Liu}}, \bibinfo {author} {\bibfnamefont {J.}~\bibnamefont {Chaloupka}},\
  and\ \bibinfo {author} {\bibfnamefont {G.}~\bibnamefont {Khaliullin}},\
  }\bibfield  {title} {\bibinfo {title} {Exchange interactions in $d^5$ kitaev
  materials: From $\mathrm{Na_2IrO_3}$ to $\alpha - \mathrm{RuCl_3}$},\ }\href
  {https://doi.org/10.1103/PhysRevB.105.214411} {\bibfield  {journal} {\bibinfo
   {journal} {Physical Review B}\ }\textbf {\bibinfo {volume} {105}},\ \bibinfo
  {pages} {214411} (\bibinfo {year} {2022})}\BibitemShut {NoStop}%
\bibitem [{\citenamefont {Liu}\ \emph {et~al.}(2011)\citenamefont {Liu},
  \citenamefont {Berlijn}, \citenamefont {Yin}, \citenamefont {Ku},
  \citenamefont {Tsvelik}, \citenamefont {Kim}, \citenamefont {Gretarsson},
  \citenamefont {Singh}, \citenamefont {Gegenwart},\ and\ \citenamefont
  {Hill}}]{Liu2011}%
  \BibitemOpen
  \bibfield  {author} {\bibinfo {author} {\bibfnamefont {X.}~\bibnamefont
  {Liu}}, \bibinfo {author} {\bibfnamefont {T.}~\bibnamefont {Berlijn}},
  \bibinfo {author} {\bibfnamefont {W.~G.}\ \bibnamefont {Yin}}, \bibinfo
  {author} {\bibfnamefont {W.}~\bibnamefont {Ku}}, \bibinfo {author}
  {\bibfnamefont {A.}~\bibnamefont {Tsvelik}}, \bibinfo {author} {\bibfnamefont
  {Y.-J.}\ \bibnamefont {Kim}}, \bibinfo {author} {\bibfnamefont
  {H.}~\bibnamefont {Gretarsson}}, \bibinfo {author} {\bibfnamefont
  {Y.}~\bibnamefont {Singh}}, \bibinfo {author} {\bibfnamefont
  {P.}~\bibnamefont {Gegenwart}},\ and\ \bibinfo {author} {\bibfnamefont
  {J.~P.}\ \bibnamefont {Hill}},\ }\bibfield  {title} {\bibinfo {title}
  {Long-range magnetic ordering in $\mathrm{Na_2IrO_3}$},\ }\href
  {https://doi.org/10.1103/PhysRevB.83.220403} {\bibfield  {journal} {\bibinfo
  {journal} {Physical Review B}\ }\textbf {\bibinfo {volume} {83}},\ \bibinfo
  {pages} {220403(R)} (\bibinfo {year} {2011})}\BibitemShut {NoStop}%
\bibitem [{\citenamefont {Ye}\ \emph {et~al.}(2012)\citenamefont {Ye},
  \citenamefont {Chi}, \citenamefont {Cao}, \citenamefont {Chakoumakos},
  \citenamefont {Fernandez-Baca}, \citenamefont {Custelcean}, \citenamefont
  {Qi}, \citenamefont {Korneta},\ and\ \citenamefont {Cao}}]{Ye2012}%
  \BibitemOpen
  \bibfield  {author} {\bibinfo {author} {\bibfnamefont {F.}~\bibnamefont
  {Ye}}, \bibinfo {author} {\bibfnamefont {S.}~\bibnamefont {Chi}}, \bibinfo
  {author} {\bibfnamefont {H.}~\bibnamefont {Cao}}, \bibinfo {author}
  {\bibfnamefont {B.~C.}\ \bibnamefont {Chakoumakos}}, \bibinfo {author}
  {\bibfnamefont {J.~A.}\ \bibnamefont {Fernandez-Baca}}, \bibinfo {author}
  {\bibfnamefont {R.}~\bibnamefont {Custelcean}}, \bibinfo {author}
  {\bibfnamefont {T.~F.}\ \bibnamefont {Qi}}, \bibinfo {author} {\bibfnamefont
  {O.~B.}\ \bibnamefont {Korneta}},\ and\ \bibinfo {author} {\bibfnamefont
  {G.}~\bibnamefont {Cao}},\ }\bibfield  {title} {\bibinfo {title} {Direct
  evidence of a zigzag spin-chain structure in the honeycomb lattice: A neutron
  and x-ray diffraction investigation of single-crystal $\mathrm{Na_2IrO_3}$},\
  }\href {https://doi.org/10.1103/PhysRevB.85.180403} {\bibfield  {journal}
  {\bibinfo  {journal} {Physical Review B}\ }\textbf {\bibinfo {volume} {85}},\
  \bibinfo {pages} {180403(R)} (\bibinfo {year} {2012})}\BibitemShut {NoStop}%
\bibitem [{\citenamefont {Biffin}\ \emph {et~al.}(2014)\citenamefont {Biffin},
  \citenamefont {Johnson}, \citenamefont {Choi}, \citenamefont {Freund},
  \citenamefont {Manni}, \citenamefont {Bombardi}, \citenamefont {Manuel},
  \citenamefont {Gegenwart},\ and\ \citenamefont {Coldea}}]{Biffin2014b}%
  \BibitemOpen
  \bibfield  {author} {\bibinfo {author} {\bibfnamefont {A.}~\bibnamefont
  {Biffin}}, \bibinfo {author} {\bibfnamefont {R.~D.}\ \bibnamefont {Johnson}},
  \bibinfo {author} {\bibfnamefont {S.}~\bibnamefont {Choi}}, \bibinfo {author}
  {\bibfnamefont {F.}~\bibnamefont {Freund}}, \bibinfo {author} {\bibfnamefont
  {S.}~\bibnamefont {Manni}}, \bibinfo {author} {\bibfnamefont
  {A.}~\bibnamefont {Bombardi}}, \bibinfo {author} {\bibfnamefont
  {P.}~\bibnamefont {Manuel}}, \bibinfo {author} {\bibfnamefont
  {P.}~\bibnamefont {Gegenwart}},\ and\ \bibinfo {author} {\bibfnamefont
  {R.}~\bibnamefont {Coldea}},\ }\bibfield  {title} {\bibinfo {title}
  {Unconventional magnetic order on the hyperhoneycomb \uppercase{K}itaev
  lattice in $\beta$-$\mathrm{Li_2IrO_3}$: Full solution via magnetic resonant
  x-ray diffraction},\ }\href {https://doi.org/10.1103/PhysRevB.90.205116}
  {\bibfield  {journal} {\bibinfo  {journal} {Physical Review B}\ }\textbf
  {\bibinfo {volume} {90}},\ \bibinfo {pages} {205116} (\bibinfo {year}
  {2014})}\BibitemShut {NoStop}%
\bibitem [{\citenamefont {{Hwan Chun}}\ \emph {et~al.}(2015)\citenamefont
  {{Hwan Chun}}, \citenamefont {Kim}, \citenamefont {Kim}, \citenamefont
  {Zheng}, \citenamefont {Stoumpos}, \citenamefont {Malliakas}, \citenamefont
  {Mitchell}, \citenamefont {Mehlawat}, \citenamefont {Singh}, \citenamefont
  {Choi}, \citenamefont {Gog}, \citenamefont {Al-Zein}, \citenamefont {Sala},
  \citenamefont {Krisch}, \citenamefont {Chaloupka}, \citenamefont {Jackeli},
  \citenamefont {Khaliullin},\ and\ \citenamefont {Kim}}]{Chun2015}%
  \BibitemOpen
  \bibfield  {author} {\bibinfo {author} {\bibfnamefont {S.}~\bibnamefont
  {{Hwan Chun}}}, \bibinfo {author} {\bibfnamefont {J.-W.}\ \bibnamefont
  {Kim}}, \bibinfo {author} {\bibfnamefont {J.}~\bibnamefont {Kim}}, \bibinfo
  {author} {\bibfnamefont {H.}~\bibnamefont {Zheng}}, \bibinfo {author}
  {\bibfnamefont {C.~C.}\ \bibnamefont {Stoumpos}}, \bibinfo {author}
  {\bibfnamefont {C.~D.}\ \bibnamefont {Malliakas}}, \bibinfo {author}
  {\bibfnamefont {J.~F.}\ \bibnamefont {Mitchell}}, \bibinfo {author}
  {\bibfnamefont {K.}~\bibnamefont {Mehlawat}}, \bibinfo {author}
  {\bibfnamefont {Y.}~\bibnamefont {Singh}}, \bibinfo {author} {\bibfnamefont
  {Y.}~\bibnamefont {Choi}}, \bibinfo {author} {\bibfnamefont {T.}~\bibnamefont
  {Gog}}, \bibinfo {author} {\bibfnamefont {A.}~\bibnamefont {Al-Zein}},
  \bibinfo {author} {\bibfnamefont {M.~M.}\ \bibnamefont {Sala}}, \bibinfo
  {author} {\bibfnamefont {M.}~\bibnamefont {Krisch}}, \bibinfo {author}
  {\bibfnamefont {J.}~\bibnamefont {Chaloupka}}, \bibinfo {author}
  {\bibfnamefont {G.}~\bibnamefont {Jackeli}}, \bibinfo {author} {\bibfnamefont
  {G.}~\bibnamefont {Khaliullin}},\ and\ \bibinfo {author} {\bibfnamefont
  {B.~J.}\ \bibnamefont {Kim}},\ }\bibfield  {title} {\bibinfo {title} {{Direct
  evidence for dominant bond-directional interactions in a honeycomb lattice
  iridate $\mathrm{Na_2IrO_3}$}},\ }\href {https://doi.org/10.1038/nphys3322}
  {\bibfield  {journal} {\bibinfo  {journal} {Nature Physics}\ }\textbf
  {\bibinfo {volume} {11}},\ \bibinfo {pages} {462} (\bibinfo {year}
  {2015})}\BibitemShut {NoStop}%
\bibitem [{\citenamefont {Williams}\ \emph {et~al.}(2016)\citenamefont
  {Williams}, \citenamefont {Johnson}, \citenamefont {Freund}, \citenamefont
  {Choi}, \citenamefont {Jesche}, \citenamefont {Kimchi}, \citenamefont
  {Manni}, \citenamefont {Bombardi}, \citenamefont {Manuel}, \citenamefont
  {Gegenwart},\ and\ \citenamefont {Coldea}}]{Williams2016}%
  \BibitemOpen
  \bibfield  {author} {\bibinfo {author} {\bibfnamefont {S.~C.}\ \bibnamefont
  {Williams}}, \bibinfo {author} {\bibfnamefont {R.~D.}\ \bibnamefont
  {Johnson}}, \bibinfo {author} {\bibfnamefont {F.}~\bibnamefont {Freund}},
  \bibinfo {author} {\bibfnamefont {S.}~\bibnamefont {Choi}}, \bibinfo {author}
  {\bibfnamefont {A.}~\bibnamefont {Jesche}}, \bibinfo {author} {\bibfnamefont
  {I.}~\bibnamefont {Kimchi}}, \bibinfo {author} {\bibfnamefont
  {S.}~\bibnamefont {Manni}}, \bibinfo {author} {\bibfnamefont
  {A.}~\bibnamefont {Bombardi}}, \bibinfo {author} {\bibfnamefont
  {P.}~\bibnamefont {Manuel}}, \bibinfo {author} {\bibfnamefont
  {P.}~\bibnamefont {Gegenwart}},\ and\ \bibinfo {author} {\bibfnamefont
  {R.}~\bibnamefont {Coldea}},\ }\bibfield  {title} {\bibinfo {title}
  {Incommensurate counterrotating magnetic order stabilized by
  \uppercase{K}itaev interactions in the layered honeycomb
  $\alpha$-$\mathrm{Li_2IrO_3}$},\ }\href
  {https://doi.org/10.1103/PhysRevB.93.195158} {\bibfield  {journal} {\bibinfo
  {journal} {Physical Review B}\ }\textbf {\bibinfo {volume} {93}},\ \bibinfo
  {pages} {195158} (\bibinfo {year} {2016})}\BibitemShut {NoStop}%
\bibitem [{\citenamefont {Abramchuk}\ \emph {et~al.}(2017)\citenamefont
  {Abramchuk}, \citenamefont {Ozsoy-Keskinbora}, \citenamefont {Krizan},
  \citenamefont {Metz}, \citenamefont {Bell},\ and\ \citenamefont
  {Tafti}}]{Abramchuk2017}%
  \BibitemOpen
  \bibfield  {author} {\bibinfo {author} {\bibfnamefont {M.}~\bibnamefont
  {Abramchuk}}, \bibinfo {author} {\bibfnamefont {C.}~\bibnamefont
  {Ozsoy-Keskinbora}}, \bibinfo {author} {\bibfnamefont {J.~W.}\ \bibnamefont
  {Krizan}}, \bibinfo {author} {\bibfnamefont {K.~R.}\ \bibnamefont {Metz}},
  \bibinfo {author} {\bibfnamefont {D.~C.}\ \bibnamefont {Bell}},\ and\
  \bibinfo {author} {\bibfnamefont {F.}~\bibnamefont {Tafti}},\ }\bibfield
  {title} {\bibinfo {title} {{\ch{Cu2IrO3}: A New Magnetically Frustrated
  Honeycomb Iridate}},\ }\href {https://doi.org/10.1021/jacs.7b06911}
  {\bibfield  {journal} {\bibinfo  {journal} {Journal of the American Chemical
  Society}\ }\textbf {\bibinfo {volume} {139}},\ \bibinfo {pages} {15371}
  (\bibinfo {year} {2017})}\BibitemShut {NoStop}%
\bibitem [{\citenamefont {Bahrami}\ \emph {et~al.}(2019)\citenamefont
  {Bahrami}, \citenamefont {Lafargue-Dit-Hauret}, \citenamefont {Lebedev},
  \citenamefont {Movshovich}, \citenamefont {Yang}, \citenamefont {Broido},
  \citenamefont {Rocquefelte},\ and\ \citenamefont {Tafti}}]{Bahrami2019}%
  \BibitemOpen
  \bibfield  {author} {\bibinfo {author} {\bibfnamefont {F.}~\bibnamefont
  {Bahrami}}, \bibinfo {author} {\bibfnamefont {W.}~\bibnamefont
  {Lafargue-Dit-Hauret}}, \bibinfo {author} {\bibfnamefont {O.~I.}\
  \bibnamefont {Lebedev}}, \bibinfo {author} {\bibfnamefont {R.}~\bibnamefont
  {Movshovich}}, \bibinfo {author} {\bibfnamefont {H.-Y.}\ \bibnamefont
  {Yang}}, \bibinfo {author} {\bibfnamefont {D.}~\bibnamefont {Broido}},
  \bibinfo {author} {\bibfnamefont {X.}~\bibnamefont {Rocquefelte}},\ and\
  \bibinfo {author} {\bibfnamefont {F.}~\bibnamefont {Tafti}},\ }\bibfield
  {title} {\bibinfo {title} {Thermodynamic evidence of proximity to a
  \uppercase{K}itaev spin liquid in $\mathrm{Ag_3LiIr_2O_6}$},\ }\href
  {https://doi.org/10.1103/PhysRevLett.123.237203} {\bibfield  {journal}
  {\bibinfo  {journal} {Physical Review Letters}\ }\textbf {\bibinfo {volume}
  {123}},\ \bibinfo {pages} {237203} (\bibinfo {year} {2019})}\BibitemShut
  {NoStop}%
\bibitem [{\citenamefont {Kitagawa}\ \emph {et~al.}(2018)\citenamefont
  {Kitagawa}, \citenamefont {Takayama}, \citenamefont {Matsumoto},
  \citenamefont {Kato}, \citenamefont {Takano}, \citenamefont {Kishimoto},
  \citenamefont {Bette}, \citenamefont {Dinnebier}, \citenamefont {Jackeli},\
  and\ \citenamefont {Takagi}}]{Kitagawa2018}%
  \BibitemOpen
  \bibfield  {author} {\bibinfo {author} {\bibfnamefont {K.}~\bibnamefont
  {Kitagawa}}, \bibinfo {author} {\bibfnamefont {T.}~\bibnamefont {Takayama}},
  \bibinfo {author} {\bibfnamefont {Y.}~\bibnamefont {Matsumoto}}, \bibinfo
  {author} {\bibfnamefont {A.}~\bibnamefont {Kato}}, \bibinfo {author}
  {\bibfnamefont {R.}~\bibnamefont {Takano}}, \bibinfo {author} {\bibfnamefont
  {Y.}~\bibnamefont {Kishimoto}}, \bibinfo {author} {\bibfnamefont
  {S.}~\bibnamefont {Bette}}, \bibinfo {author} {\bibfnamefont
  {R.}~\bibnamefont {Dinnebier}}, \bibinfo {author} {\bibfnamefont
  {G.}~\bibnamefont {Jackeli}},\ and\ \bibinfo {author} {\bibfnamefont
  {H.}~\bibnamefont {Takagi}},\ }\bibfield  {title} {\bibinfo {title} {A
  spin-orbital-entangled quantum liquid on a honeycomb lattice},\ }\href
  {https://doi.org/10.1038/nature25482} {\bibfield  {journal} {\bibinfo
  {journal} {Nature}\ }\textbf {\bibinfo {volume} {554}},\ \bibinfo {pages}
  {341} (\bibinfo {year} {2018})}\BibitemShut {NoStop}%
\bibitem [{\citenamefont {Kenney}\ \emph {et~al.}(2019)\citenamefont {Kenney},
  \citenamefont {Segre}, \citenamefont {Lafargue-Dit-Hauret}, \citenamefont
  {Lebedev}, \citenamefont {Abramchuk}, \citenamefont {Berlie}, \citenamefont
  {Cottrell}, \citenamefont {Simutis}, \citenamefont {Bahrami}, \citenamefont
  {Mordvinova}, \citenamefont {Fabbris}, \citenamefont {McChesney},
  \citenamefont {Haskel}, \citenamefont {Rocquefelte}, \citenamefont {Graf},\
  and\ \citenamefont {Tafti}}]{Kenney2019}%
  \BibitemOpen
  \bibfield  {author} {\bibinfo {author} {\bibfnamefont {E.~M.}\ \bibnamefont
  {Kenney}}, \bibinfo {author} {\bibfnamefont {C.~U.}\ \bibnamefont {Segre}},
  \bibinfo {author} {\bibfnamefont {W.}~\bibnamefont {Lafargue-Dit-Hauret}},
  \bibinfo {author} {\bibfnamefont {O.~I.}\ \bibnamefont {Lebedev}}, \bibinfo
  {author} {\bibfnamefont {M.}~\bibnamefont {Abramchuk}}, \bibinfo {author}
  {\bibfnamefont {A.}~\bibnamefont {Berlie}}, \bibinfo {author} {\bibfnamefont
  {S.~P.}\ \bibnamefont {Cottrell}}, \bibinfo {author} {\bibfnamefont
  {G.}~\bibnamefont {Simutis}}, \bibinfo {author} {\bibfnamefont
  {F.}~\bibnamefont {Bahrami}}, \bibinfo {author} {\bibfnamefont {N.~E.}\
  \bibnamefont {Mordvinova}}, \bibinfo {author} {\bibfnamefont
  {G.}~\bibnamefont {Fabbris}}, \bibinfo {author} {\bibfnamefont {J.~L.}\
  \bibnamefont {McChesney}}, \bibinfo {author} {\bibfnamefont {D.}~\bibnamefont
  {Haskel}}, \bibinfo {author} {\bibfnamefont {X.}~\bibnamefont {Rocquefelte}},
  \bibinfo {author} {\bibfnamefont {M.~J.}\ \bibnamefont {Graf}},\ and\
  \bibinfo {author} {\bibfnamefont {F.}~\bibnamefont {Tafti}},\ }\bibfield
  {title} {\bibinfo {title} {{Coexistence of static and dynamic magnetism in
  the \uppercase{K}itaev spin liquid material \ch{Cu2IrO3}}},\ }\href
  {https://doi.org/10.1103/PhysRevB.100.094418} {\bibfield  {journal} {\bibinfo
   {journal} {Physical Review B}\ }\textbf {\bibinfo {volume} {100}},\ \bibinfo
  {pages} {094418} (\bibinfo {year} {2019})}\BibitemShut {NoStop}%
\bibitem [{\citenamefont {Knolle}\ \emph {et~al.}(2019)\citenamefont {Knolle},
  \citenamefont {Moessner},\ and\ \citenamefont {Perkins}}]{Knolle2019}%
  \BibitemOpen
  \bibfield  {author} {\bibinfo {author} {\bibfnamefont {J.}~\bibnamefont
  {Knolle}}, \bibinfo {author} {\bibfnamefont {R.}~\bibnamefont {Moessner}},\
  and\ \bibinfo {author} {\bibfnamefont {N.~B.}\ \bibnamefont {Perkins}},\
  }\bibfield  {title} {\bibinfo {title} {Bond-disordered spin liquid and the
  honeycomb iridate $\mathrm{H_3LiIr_2O_6}$: Abundant low-energy density of
  states from random majorana hopping},\ }\href
  {https://doi.org/10.1103/PhysRevLett.122.047202} {\bibfield  {journal}
  {\bibinfo  {journal} {Physical Review Letters}\ }\textbf {\bibinfo {volume}
  {122}},\ \bibinfo {pages} {047202} (\bibinfo {year} {2019})}\BibitemShut
  {NoStop}%
\bibitem [{\citenamefont {de~la Torre}\ \emph {et~al.}(2021)\citenamefont
  {de~la Torre}, \citenamefont {Zager}, \citenamefont {Bahrami}, \citenamefont
  {DiScala}, \citenamefont {Chamorro}, \citenamefont {Upton}, \citenamefont
  {Fabbris}, \citenamefont {Haskel}, \citenamefont {Casa}, \citenamefont
  {McQueen}, \citenamefont {Tafti},\ and\ \citenamefont
  {Plumb}}]{delaTorre2021}%
  \BibitemOpen
  \bibfield  {author} {\bibinfo {author} {\bibfnamefont {A.}~\bibnamefont
  {de~la Torre}}, \bibinfo {author} {\bibfnamefont {B.}~\bibnamefont {Zager}},
  \bibinfo {author} {\bibfnamefont {F.}~\bibnamefont {Bahrami}}, \bibinfo
  {author} {\bibfnamefont {M.}~\bibnamefont {DiScala}}, \bibinfo {author}
  {\bibfnamefont {J.~R.}\ \bibnamefont {Chamorro}}, \bibinfo {author}
  {\bibfnamefont {M.~H.}\ \bibnamefont {Upton}}, \bibinfo {author}
  {\bibfnamefont {G.}~\bibnamefont {Fabbris}}, \bibinfo {author} {\bibfnamefont
  {D.}~\bibnamefont {Haskel}}, \bibinfo {author} {\bibfnamefont
  {D.}~\bibnamefont {Casa}}, \bibinfo {author} {\bibfnamefont {T.~M.}\
  \bibnamefont {McQueen}}, \bibinfo {author} {\bibfnamefont {F.}~\bibnamefont
  {Tafti}},\ and\ \bibinfo {author} {\bibfnamefont {K.~W.}\ \bibnamefont
  {Plumb}},\ }\bibfield  {title} {\bibinfo {title} {Enhanced hybridization in
  the electronic ground state of the intercalated honeycomb iridate
  $\mathrm{Ag_3LiIr_2O_6}$},\ }\href
  {https://journals.aps.org/prb/abstract/10.1103/PhysRevB.104.L100416}
  {\bibfield  {journal} {\bibinfo  {journal} {Physical Review B}\ }\textbf
  {\bibinfo {volume} {104}},\ \bibinfo {pages} {L100416} (\bibinfo {year}
  {2021})}\BibitemShut {NoStop}%
\bibitem [{\citenamefont {Bahrami}\ \emph {et~al.}(2022)\citenamefont
  {Bahrami}, \citenamefont {Abramchuk}, \citenamefont {Lebedev},\ and\
  \citenamefont {Tafti}}]{Bahrami2022}%
  \BibitemOpen
  \bibfield  {author} {\bibinfo {author} {\bibfnamefont {F.}~\bibnamefont
  {Bahrami}}, \bibinfo {author} {\bibfnamefont {M.}~\bibnamefont {Abramchuk}},
  \bibinfo {author} {\bibfnamefont {O.}~\bibnamefont {Lebedev}},\ and\ \bibinfo
  {author} {\bibfnamefont {F.}~\bibnamefont {Tafti}},\ }\bibfield  {title}
  {\bibinfo {title} {Metastable \uppercase{K}itaev magnets},\ }\href
  {https://doi.org/10.3390/molecules27030871} {\bibfield  {journal} {\bibinfo
  {journal} {Molecules}\ }\textbf {\bibinfo {volume} {27}},\ \bibinfo {pages}
  {871} (\bibinfo {year} {2022})}\BibitemShut {NoStop}%
\bibitem [{\citenamefont {Fabbris}\ \emph {et~al.}(2021)\citenamefont
  {Fabbris}, \citenamefont {Thorn}, \citenamefont {Bi}, \citenamefont
  {Abramchuk}, \citenamefont {Bahrami}, \citenamefont {Kim}, \citenamefont
  {Shinmei}, \citenamefont {Irifune}, \citenamefont {Tafti}, \citenamefont
  {Kolmogorov},\ and\ \citenamefont {Haskel}}]{Fabbris2021}%
  \BibitemOpen
  \bibfield  {author} {\bibinfo {author} {\bibfnamefont {G.}~\bibnamefont
  {Fabbris}}, \bibinfo {author} {\bibfnamefont {A.}~\bibnamefont {Thorn}},
  \bibinfo {author} {\bibfnamefont {W.}~\bibnamefont {Bi}}, \bibinfo {author}
  {\bibfnamefont {M.}~\bibnamefont {Abramchuk}}, \bibinfo {author}
  {\bibfnamefont {F.}~\bibnamefont {Bahrami}}, \bibinfo {author} {\bibfnamefont
  {J.~H.}\ \bibnamefont {Kim}}, \bibinfo {author} {\bibfnamefont
  {T.}~\bibnamefont {Shinmei}}, \bibinfo {author} {\bibfnamefont
  {T.}~\bibnamefont {Irifune}}, \bibinfo {author} {\bibfnamefont
  {F.}~\bibnamefont {Tafti}}, \bibinfo {author} {\bibfnamefont {A.~N.}\
  \bibnamefont {Kolmogorov}},\ and\ \bibinfo {author} {\bibfnamefont
  {D.}~\bibnamefont {Haskel}},\ }\bibfield  {title} {\bibinfo {title} {Complex
  pressure-temperature structural phase diagram of the honeycomb iridate
  $\mathrm{Cu_2IrO_3}$},\ }\href {https://doi.org/10.1103/PhysRevB.104.014102}
  {\bibfield  {journal} {\bibinfo  {journal} {Physical Review B}\ }\textbf
  {\bibinfo {volume} {104}},\ \bibinfo {pages} {014102} (\bibinfo {year}
  {2021})}\BibitemShut {NoStop}%
\bibitem [{\citenamefont {Jin}\ \emph {et~al.}(2022)\citenamefont {Jin},
  \citenamefont {Wang}, \citenamefont {Jin}, \citenamefont {Jiang},
  \citenamefont {Jiang}, \citenamefont {Li}, \citenamefont {Nakamoto},
  \citenamefont {Shimizu},\ and\ \citenamefont {Zhu}}]{Jin2022}%
  \BibitemOpen
  \bibfield  {author} {\bibinfo {author} {\bibfnamefont {C.}~\bibnamefont
  {Jin}}, \bibinfo {author} {\bibfnamefont {Y.}~\bibnamefont {Wang}}, \bibinfo
  {author} {\bibfnamefont {M.}~\bibnamefont {Jin}}, \bibinfo {author}
  {\bibfnamefont {Z.}~\bibnamefont {Jiang}}, \bibinfo {author} {\bibfnamefont
  {D.}~\bibnamefont {Jiang}}, \bibinfo {author} {\bibfnamefont
  {J.}~\bibnamefont {Li}}, \bibinfo {author} {\bibfnamefont {Y.}~\bibnamefont
  {Nakamoto}}, \bibinfo {author} {\bibfnamefont {K.}~\bibnamefont {Shimizu}},\
  and\ \bibinfo {author} {\bibfnamefont {J.}~\bibnamefont {Zhu}},\ }\bibfield
  {title} {\bibinfo {title} {Insulator-metal transition and crossover from
  negative to positive magnetoresistance in $\mathrm{Cu_2IrO_3}$ under high
  pressure},\ }\href
  {https://journals.aps.org/prb/abstract/10.1103/PhysRevB.105.144402}
  {\bibfield  {journal} {\bibinfo  {journal} {Physical Review B}\ }\textbf
  {\bibinfo {volume} {105}},\ \bibinfo {pages} {144402} (\bibinfo {year}
  {2022})}\BibitemShut {NoStop}%
\bibitem [{\citenamefont {Pal}\ \emph {et~al.}(2023)\citenamefont {Pal},
  \citenamefont {Malavi}, \citenamefont {Sinha}, \citenamefont {Ali},
  \citenamefont {Sakrikar}, \citenamefont {Joseph}, \citenamefont {Waghmare},
  \citenamefont {Singh}, \citenamefont {Muthu}, \citenamefont {Karmakar},\ and\
  \citenamefont {Sood}}]{Pal2023}%
  \BibitemOpen
  \bibfield  {author} {\bibinfo {author} {\bibfnamefont {S.}~\bibnamefont
  {Pal}}, \bibinfo {author} {\bibfnamefont {P.}~\bibnamefont {Malavi}},
  \bibinfo {author} {\bibfnamefont {A.}~\bibnamefont {Sinha}}, \bibinfo
  {author} {\bibfnamefont {A.}~\bibnamefont {Ali}}, \bibinfo {author}
  {\bibfnamefont {P.}~\bibnamefont {Sakrikar}}, \bibinfo {author}
  {\bibfnamefont {B.}~\bibnamefont {Joseph}}, \bibinfo {author} {\bibfnamefont
  {U.~V.}\ \bibnamefont {Waghmare}}, \bibinfo {author} {\bibfnamefont
  {Y.}~\bibnamefont {Singh}}, \bibinfo {author} {\bibfnamefont {D.~V.~S.}\
  \bibnamefont {Muthu}}, \bibinfo {author} {\bibfnamefont {S.}~\bibnamefont
  {Karmakar}},\ and\ \bibinfo {author} {\bibfnamefont {A.~K.}\ \bibnamefont
  {Sood}},\ }\bibfield  {title} {\bibinfo {title} {Pressure tuning of
  structure, magnetic frustration, and carrier conduction in the
  \uppercase{K}itaev spin liquid candidate $\mathrm{Cu_2IrO_3}$},\ }\href
  {https://journals.aps.org/prb/abstract/10.1103/PhysRevB.107.085105}
  {\bibfield  {journal} {\bibinfo  {journal} {Physical Review B}\ }\textbf
  {\bibinfo {volume} {107}},\ \bibinfo {pages} {085105} (\bibinfo {year}
  {2023})}\BibitemShut {NoStop}%
\bibitem [{\citenamefont {Veiga}\ \emph {et~al.}(2017)\citenamefont {Veiga},
  \citenamefont {Etter}, \citenamefont {Glazyrin}, \citenamefont {Sun},
  \citenamefont {Escanhoela}, \citenamefont {Fabbris}, \citenamefont
  {Mardegan}, \citenamefont {Malavi}, \citenamefont {Deng}, \citenamefont
  {Stavropoulos}, \citenamefont {Kee}, \citenamefont {Yang}, \citenamefont
  {{Van Veenendaal}}, \citenamefont {Schilling}, \citenamefont {Takayama},
  \citenamefont {Takagi},\ and\ \citenamefont {Haskel}}]{Veiga2017}%
  \BibitemOpen
  \bibfield  {author} {\bibinfo {author} {\bibfnamefont {L.~S.~I.}\
  \bibnamefont {Veiga}}, \bibinfo {author} {\bibfnamefont {M.}~\bibnamefont
  {Etter}}, \bibinfo {author} {\bibfnamefont {K.}~\bibnamefont {Glazyrin}},
  \bibinfo {author} {\bibfnamefont {F.}~\bibnamefont {Sun}}, \bibinfo {author}
  {\bibfnamefont {C.~A.}\ \bibnamefont {Escanhoela}}, \bibinfo {author}
  {\bibfnamefont {G.}~\bibnamefont {Fabbris}}, \bibinfo {author} {\bibfnamefont
  {J.~R.~L.}\ \bibnamefont {Mardegan}}, \bibinfo {author} {\bibfnamefont
  {P.~S.}\ \bibnamefont {Malavi}}, \bibinfo {author} {\bibfnamefont
  {Y.}~\bibnamefont {Deng}}, \bibinfo {author} {\bibfnamefont {P.~P.}\
  \bibnamefont {Stavropoulos}}, \bibinfo {author} {\bibfnamefont {H.~Y.}\
  \bibnamefont {Kee}}, \bibinfo {author} {\bibfnamefont {W.~G.}\ \bibnamefont
  {Yang}}, \bibinfo {author} {\bibfnamefont {M.}~\bibnamefont {{Van
  Veenendaal}}}, \bibinfo {author} {\bibfnamefont {J.~S.}\ \bibnamefont
  {Schilling}}, \bibinfo {author} {\bibfnamefont {T.}~\bibnamefont {Takayama}},
  \bibinfo {author} {\bibfnamefont {H.}~\bibnamefont {Takagi}},\ and\ \bibinfo
  {author} {\bibfnamefont {D.}~\bibnamefont {Haskel}},\ }\bibfield  {title}
  {\bibinfo {title} {{Pressure tuning of bond-directional exchange interactions
  and magnetic frustration in the hyperhoneycomb iridate
  $\beta\mathrm{-Li_2IrO_3}$}},\ }\href
  {https://doi.org/10.1103/PhysRevB.96.140402} {\bibfield  {journal} {\bibinfo
  {journal} {Physical Review B}\ }\textbf {\bibinfo {volume} {96}},\ \bibinfo
  {pages} {140402(R)} (\bibinfo {year} {2017})}\BibitemShut {NoStop}%
\bibitem [{\citenamefont {Veiga}\ \emph {et~al.}(2019)\citenamefont {Veiga},
  \citenamefont {Glazyrin}, \citenamefont {Fabbris}, \citenamefont {Dashwood},
  \citenamefont {Vale}, \citenamefont {Park}, \citenamefont {Etter},
  \citenamefont {Irifune}, \citenamefont {Pascarelli}, \citenamefont
  {McMorrow}, \citenamefont {Takayama}, \citenamefont {Takagi},\ and\
  \citenamefont {Haskel}}]{Veiga2019}%
  \BibitemOpen
  \bibfield  {author} {\bibinfo {author} {\bibfnamefont {L.~S.~I.}\
  \bibnamefont {Veiga}}, \bibinfo {author} {\bibfnamefont {K.}~\bibnamefont
  {Glazyrin}}, \bibinfo {author} {\bibfnamefont {G.}~\bibnamefont {Fabbris}},
  \bibinfo {author} {\bibfnamefont {C.~D.}\ \bibnamefont {Dashwood}}, \bibinfo
  {author} {\bibfnamefont {J.~G.}\ \bibnamefont {Vale}}, \bibinfo {author}
  {\bibfnamefont {H.}~\bibnamefont {Park}}, \bibinfo {author} {\bibfnamefont
  {M.}~\bibnamefont {Etter}}, \bibinfo {author} {\bibfnamefont
  {T.}~\bibnamefont {Irifune}}, \bibinfo {author} {\bibfnamefont
  {S.}~\bibnamefont {Pascarelli}}, \bibinfo {author} {\bibfnamefont {D.~F.}\
  \bibnamefont {McMorrow}}, \bibinfo {author} {\bibfnamefont {T.}~\bibnamefont
  {Takayama}}, \bibinfo {author} {\bibfnamefont {H.}~\bibnamefont {Takagi}},\
  and\ \bibinfo {author} {\bibfnamefont {D.}~\bibnamefont {Haskel}},\
  }\bibfield  {title} {\bibinfo {title} {{Pressure-induced structural
  dimerization in the hyperhoneycomb iridate \ch{$\beta$-Li2IrO3} at low
  temperatures}},\ }\href {https://doi.org/10.1103/PhysRevB.100.064104}
  {\bibfield  {journal} {\bibinfo  {journal} {Physical Review B}\ }\textbf
  {\bibinfo {volume} {100}},\ \bibinfo {pages} {064104} (\bibinfo {year}
  {2019})}\BibitemShut {NoStop}%
\bibitem [{\citenamefont {Hermann}\ \emph {et~al.}(2018)\citenamefont
  {Hermann}, \citenamefont {Altmeyer}, \citenamefont {Ebad-Allah},
  \citenamefont {Freund}, \citenamefont {Jesche}, \citenamefont {Tsirlin},
  \citenamefont {Hanfland}, \citenamefont {Gegenwart}, \citenamefont {Mazin},
  \citenamefont {Khomskii}, \citenamefont {Valent{\'{i}}},\ and\ \citenamefont
  {Kuntscher}}]{Hermann2018}%
  \BibitemOpen
  \bibfield  {author} {\bibinfo {author} {\bibfnamefont {V.}~\bibnamefont
  {Hermann}}, \bibinfo {author} {\bibfnamefont {M.}~\bibnamefont {Altmeyer}},
  \bibinfo {author} {\bibfnamefont {J.}~\bibnamefont {Ebad-Allah}}, \bibinfo
  {author} {\bibfnamefont {F.}~\bibnamefont {Freund}}, \bibinfo {author}
  {\bibfnamefont {A.}~\bibnamefont {Jesche}}, \bibinfo {author} {\bibfnamefont
  {A.~A.}\ \bibnamefont {Tsirlin}}, \bibinfo {author} {\bibfnamefont
  {M.}~\bibnamefont {Hanfland}}, \bibinfo {author} {\bibfnamefont
  {P.}~\bibnamefont {Gegenwart}}, \bibinfo {author} {\bibfnamefont {I.~I.}\
  \bibnamefont {Mazin}}, \bibinfo {author} {\bibfnamefont {D.~I.}\ \bibnamefont
  {Khomskii}}, \bibinfo {author} {\bibfnamefont {R.}~\bibnamefont
  {Valent{\'{i}}}},\ and\ \bibinfo {author} {\bibfnamefont {C.~A.}\
  \bibnamefont {Kuntscher}},\ }\bibfield  {title} {\bibinfo {title}
  {{Competition between spin-orbit coupling, magnetism, and dimerization in the
  honeycomb iridates: $\mathrm{\alpha-Li_2IrO_3}$ under pressure}},\ }\href
  {https://link.aps.org/doi/10.1103/PhysRevB.97.020104} {\bibfield  {journal}
  {\bibinfo  {journal} {Physical Review B}\ }\textbf {\bibinfo {volume} {97}},\
  \bibinfo {pages} {020104(R)} (\bibinfo {year} {2018})}\BibitemShut {NoStop}%
\bibitem [{\citenamefont {Clancy}\ \emph {et~al.}(2018)\citenamefont {Clancy},
  \citenamefont {Gretarsson}, \citenamefont {Sears}, \citenamefont {Singh},
  \citenamefont {Desgreniers}, \citenamefont {Mehlawat}, \citenamefont {Layek},
  \citenamefont {Rozenberg}, \citenamefont {Ding}, \citenamefont {Upton},
  \citenamefont {Casa}, \citenamefont {Chen}, \citenamefont {Im}, \citenamefont
  {Lee}, \citenamefont {Yadav}, \citenamefont {Hozoi}, \citenamefont {Efremov},
  \citenamefont {van~den Brink},\ and\ \citenamefont {Kim}}]{Clancy2018}%
  \BibitemOpen
  \bibfield  {author} {\bibinfo {author} {\bibfnamefont {J.~P.}\ \bibnamefont
  {Clancy}}, \bibinfo {author} {\bibfnamefont {H.}~\bibnamefont {Gretarsson}},
  \bibinfo {author} {\bibfnamefont {J.~A.}\ \bibnamefont {Sears}}, \bibinfo
  {author} {\bibfnamefont {Y.}~\bibnamefont {Singh}}, \bibinfo {author}
  {\bibfnamefont {S.}~\bibnamefont {Desgreniers}}, \bibinfo {author}
  {\bibfnamefont {K.}~\bibnamefont {Mehlawat}}, \bibinfo {author}
  {\bibfnamefont {S.}~\bibnamefont {Layek}}, \bibinfo {author} {\bibfnamefont
  {G.~K.}\ \bibnamefont {Rozenberg}}, \bibinfo {author} {\bibfnamefont
  {Y.}~\bibnamefont {Ding}}, \bibinfo {author} {\bibfnamefont {M.~H.}\
  \bibnamefont {Upton}}, \bibinfo {author} {\bibfnamefont {D.}~\bibnamefont
  {Casa}}, \bibinfo {author} {\bibfnamefont {N.}~\bibnamefont {Chen}}, \bibinfo
  {author} {\bibfnamefont {J.}~\bibnamefont {Im}}, \bibinfo {author}
  {\bibfnamefont {Y.}~\bibnamefont {Lee}}, \bibinfo {author} {\bibfnamefont
  {R.}~\bibnamefont {Yadav}}, \bibinfo {author} {\bibfnamefont
  {L.}~\bibnamefont {Hozoi}}, \bibinfo {author} {\bibfnamefont
  {D.}~\bibnamefont {Efremov}}, \bibinfo {author} {\bibfnamefont
  {J.}~\bibnamefont {van~den Brink}},\ and\ \bibinfo {author} {\bibfnamefont
  {Y.-J.}\ \bibnamefont {Kim}},\ }\bibfield  {title} {\bibinfo {title}
  {{Pressure-driven collapse of the relativistic electronic ground state in a
  honeycomb iridate}},\ }\href {https://doi.org/10.1038/s41535-018-0109-0}
  {\bibfield  {journal} {\bibinfo  {journal} {npj Quantum Materials}\ }\textbf
  {\bibinfo {volume} {3}},\ \bibinfo {pages} {35} (\bibinfo {year}
  {2018})}\BibitemShut {NoStop}%
\bibitem [{\citenamefont {Jin}\ \emph {et~al.}(2024)\citenamefont {Jin},
  \citenamefont {Han}, \citenamefont {Zheng}, \citenamefont {Chen},
  \citenamefont {Pei}, \citenamefont {Nakamoto}, \citenamefont {Shimizu},
  \citenamefont {Wang},\ and\ \citenamefont {Zhu}}]{Jin2024}%
  \BibitemOpen
  \bibfield  {author} {\bibinfo {author} {\bibfnamefont {C.}~\bibnamefont
  {Jin}}, \bibinfo {author} {\bibfnamefont {J.}~\bibnamefont {Han}}, \bibinfo
  {author} {\bibfnamefont {Q.}~\bibnamefont {Zheng}}, \bibinfo {author}
  {\bibfnamefont {E.}~\bibnamefont {Chen}}, \bibinfo {author} {\bibfnamefont
  {T.}~\bibnamefont {Pei}}, \bibinfo {author} {\bibfnamefont {Y.}~\bibnamefont
  {Nakamoto}}, \bibinfo {author} {\bibfnamefont {K.}~\bibnamefont {Shimizu}},
  \bibinfo {author} {\bibfnamefont {Y.}~\bibnamefont {Wang}},\ and\ \bibinfo
  {author} {\bibfnamefont {J.}~\bibnamefont {Zhu}},\ }\bibfield  {title}
  {\bibinfo {title} {Pressure-induced dimerization and crossover from negative
  to positive magnetoresistance in $\mathrm{Ag_3LiIr_2O_6}$},\ }\href
  {https://doi.org/10.1103/PhysRevB.109.094411} {\bibfield  {journal} {\bibinfo
   {journal} {Physical Review B}\ }\textbf {\bibinfo {volume} {109}},\ \bibinfo
  {pages} {094411} (\bibinfo {year} {2024})}\BibitemShut {NoStop}%
\bibitem [{\citenamefont {Takayama}\ \emph {et~al.}(2019)\citenamefont
  {Takayama}, \citenamefont {Krajewska}, \citenamefont {Gibbs}, \citenamefont
  {Yaresko}, \citenamefont {Ishii}, \citenamefont {Yamaoka}, \citenamefont
  {Ishii}, \citenamefont {Hiraoka}, \citenamefont {Funnell}, \citenamefont
  {Bull},\ and\ \citenamefont {Takagi}}]{Takayama2019}%
  \BibitemOpen
  \bibfield  {author} {\bibinfo {author} {\bibfnamefont {T.}~\bibnamefont
  {Takayama}}, \bibinfo {author} {\bibfnamefont {A.}~\bibnamefont {Krajewska}},
  \bibinfo {author} {\bibfnamefont {A.~S.}\ \bibnamefont {Gibbs}}, \bibinfo
  {author} {\bibfnamefont {A.~N.}\ \bibnamefont {Yaresko}}, \bibinfo {author}
  {\bibfnamefont {H.}~\bibnamefont {Ishii}}, \bibinfo {author} {\bibfnamefont
  {H.}~\bibnamefont {Yamaoka}}, \bibinfo {author} {\bibfnamefont
  {K.}~\bibnamefont {Ishii}}, \bibinfo {author} {\bibfnamefont
  {N.}~\bibnamefont {Hiraoka}}, \bibinfo {author} {\bibfnamefont {N.~P.}\
  \bibnamefont {Funnell}}, \bibinfo {author} {\bibfnamefont {C.~L.}\
  \bibnamefont {Bull}},\ and\ \bibinfo {author} {\bibfnamefont
  {H.}~\bibnamefont {Takagi}},\ }\bibfield  {title} {\bibinfo {title}
  {{Pressure-induced collapse of the spin-orbital Mott state in the
  hyperhoneycomb iridate \ch{$\beta$-Li2IrO3}}},\ }\href
  {https://doi.org/10.1103/PhysRevB.99.125127} {\bibfield  {journal} {\bibinfo
  {journal} {Physical Review B}\ }\textbf {\bibinfo {volume} {99}},\ \bibinfo
  {pages} {125127} (\bibinfo {year} {2019})}\BibitemShut {NoStop}%
\bibitem [{\citenamefont {Thole}\ and\ \citenamefont {van~der
  Laan}(1988)}]{Thole1988}%
  \BibitemOpen
  \bibfield  {author} {\bibinfo {author} {\bibfnamefont {B.~T.}\ \bibnamefont
  {Thole}}\ and\ \bibinfo {author} {\bibfnamefont {G.}~\bibnamefont {van~der
  Laan}},\ }\bibfield  {title} {\bibinfo {title} {Linear relation between x-ray
  absorption branching ratio and valence-band spin-orbit expectation value},\
  }\href {https://doi.org/10.1103/PhysRevA.38.1943} {\bibfield  {journal}
  {\bibinfo  {journal} {Physical Review A}\ }\textbf {\bibinfo {volume} {38}},\
  \bibinfo {pages} {1943} (\bibinfo {year} {1988})}\BibitemShut {NoStop}%
\bibitem [{\citenamefont {Carra}\ \emph {et~al.}(1993)\citenamefont {Carra},
  \citenamefont {Thole}, \citenamefont {Altarelli},\ and\ \citenamefont
  {Wang}}]{Carra1993}%
  \BibitemOpen
  \bibfield  {author} {\bibinfo {author} {\bibfnamefont {P.}~\bibnamefont
  {Carra}}, \bibinfo {author} {\bibfnamefont {B.~T.}\ \bibnamefont {Thole}},
  \bibinfo {author} {\bibfnamefont {M.}~\bibnamefont {Altarelli}},\ and\
  \bibinfo {author} {\bibfnamefont {X.}~\bibnamefont {Wang}},\ }\bibfield
  {title} {\bibinfo {title} {X-ray circular dichroism and local magnetic
  fields},\ }\href {https://doi.org/10.1103/PhysRevLett.70.694} {\bibfield
  {journal} {\bibinfo  {journal} {Physical Review Letters}\ }\textbf {\bibinfo
  {volume} {70}},\ \bibinfo {pages} {694} (\bibinfo {year} {1993})}\BibitemShut
  {NoStop}%
\bibitem [{Com()}]{CommentTz}%
  \BibitemOpen
  \href@noop {} {}\bibinfo {note} {The spin sum rule depends on the magnetic
  dipole operator $\langle T_z \rangle$ \cite{Carra1993}. We use $\langle T_z
  \rangle / \langle S_z \rangle = 0.18$ obtained from theoretical calculations
  of $\mathrm{Sr_2IrO_4}$ \cite{Haskel2012}.}\BibitemShut {Stop}%
\bibitem [{\citenamefont {Clancy}\ \emph {et~al.}(2012)\citenamefont {Clancy},
  \citenamefont {Chen}, \citenamefont {Kim}, \citenamefont {Chen},
  \citenamefont {Plumb}, \citenamefont {Jeon}, \citenamefont {Noh},\ and\
  \citenamefont {Kim}}]{Clancy2012}%
  \BibitemOpen
  \bibfield  {author} {\bibinfo {author} {\bibfnamefont {J.~P.}\ \bibnamefont
  {Clancy}}, \bibinfo {author} {\bibfnamefont {N.}~\bibnamefont {Chen}},
  \bibinfo {author} {\bibfnamefont {C.~Y.}\ \bibnamefont {Kim}}, \bibinfo
  {author} {\bibfnamefont {W.~F.}\ \bibnamefont {Chen}}, \bibinfo {author}
  {\bibfnamefont {K.~W.}\ \bibnamefont {Plumb}}, \bibinfo {author}
  {\bibfnamefont {B.~C.}\ \bibnamefont {Jeon}}, \bibinfo {author}
  {\bibfnamefont {T.~W.}\ \bibnamefont {Noh}},\ and\ \bibinfo {author}
  {\bibfnamefont {Y.-J.}\ \bibnamefont {Kim}},\ }\bibfield  {title} {\bibinfo
  {title} {{Spin-orbit coupling in iridium-based 5d compounds probed by x-ray
  absorption spectroscopy}},\ }\href
  {https://doi.org/10.1103/PhysRevB.86.195131} {\bibfield  {journal} {\bibinfo
  {journal} {Physical Review B}\ }\textbf {\bibinfo {volume} {86}},\ \bibinfo
  {pages} {195131} (\bibinfo {year} {2012})}\BibitemShut {NoStop}%
\bibitem [{\citenamefont {Laguna-Marco}\ \emph {et~al.}(2010)\citenamefont
  {Laguna-Marco}, \citenamefont {Haskel}, \citenamefont {Souza-Neto},
  \citenamefont {Lang}, \citenamefont {Krishnamurthy}, \citenamefont {Chikara},
  \citenamefont {Cao},\ and\ \citenamefont {van
  Veenendaal}}]{Laguna-Marco2010}%
  \BibitemOpen
  \bibfield  {author} {\bibinfo {author} {\bibfnamefont {M.~A.}\ \bibnamefont
  {Laguna-Marco}}, \bibinfo {author} {\bibfnamefont {D.}~\bibnamefont
  {Haskel}}, \bibinfo {author} {\bibfnamefont {N.}~\bibnamefont {Souza-Neto}},
  \bibinfo {author} {\bibfnamefont {J.~C.}\ \bibnamefont {Lang}}, \bibinfo
  {author} {\bibfnamefont {V.~V.}\ \bibnamefont {Krishnamurthy}}, \bibinfo
  {author} {\bibfnamefont {S.}~\bibnamefont {Chikara}}, \bibinfo {author}
  {\bibfnamefont {G.}~\bibnamefont {Cao}},\ and\ \bibinfo {author}
  {\bibfnamefont {M.}~\bibnamefont {van Veenendaal}},\ }\bibfield  {title}
  {\bibinfo {title} {{Orbital Magnetism and Spin-Orbit Effects in the
  Electronic Structure of $\mathrm{BaIrO_3}$}},\ }\href
  {https://doi.org/10.1103/PhysRevLett.105.216407} {\bibfield  {journal}
  {\bibinfo  {journal} {Physical Review Letters}\ }\textbf {\bibinfo {volume}
  {105}},\ \bibinfo {pages} {216407} (\bibinfo {year} {2010})}\BibitemShut
  {NoStop}%
\bibitem [{\citenamefont {Haskel}\ \emph {et~al.}(2012)\citenamefont {Haskel},
  \citenamefont {Fabbris}, \citenamefont {Zhernenkov}, \citenamefont {Kong},
  \citenamefont {Jin}, \citenamefont {Cao},\ and\ \citenamefont {van
  Veenendaal}}]{Haskel2012}%
  \BibitemOpen
  \bibfield  {author} {\bibinfo {author} {\bibfnamefont {D.}~\bibnamefont
  {Haskel}}, \bibinfo {author} {\bibfnamefont {G.}~\bibnamefont {Fabbris}},
  \bibinfo {author} {\bibfnamefont {M.}~\bibnamefont {Zhernenkov}}, \bibinfo
  {author} {\bibfnamefont {P.~P.}\ \bibnamefont {Kong}}, \bibinfo {author}
  {\bibfnamefont {C.~Q.}\ \bibnamefont {Jin}}, \bibinfo {author} {\bibfnamefont
  {G.}~\bibnamefont {Cao}},\ and\ \bibinfo {author} {\bibfnamefont
  {M.}~\bibnamefont {van Veenendaal}},\ }\bibfield  {title} {\bibinfo {title}
  {{Pressure Tuning of the Spin-Orbit Coupled Ground State in
  $\mathrm{Sr_2IrO_4}$}},\ }\href
  {https://doi.org/10.1103/PhysRevLett.109.027204} {\bibfield  {journal}
  {\bibinfo  {journal} {Physical Review Letters}\ }\textbf {\bibinfo {volume}
  {109}},\ \bibinfo {pages} {027204} (\bibinfo {year} {2012})}\BibitemShut
  {NoStop}%
\bibitem [{\citenamefont {Laguna-Marco}\ \emph {et~al.}(2015)\citenamefont
  {Laguna-Marco}, \citenamefont {Kayser}, \citenamefont {Alonso}, \citenamefont
  {Mart{\'{i}}nez-Lope}, \citenamefont {van Veenendaal}, \citenamefont {Choi},\
  and\ \citenamefont {Haskel}}]{Laguna-Marco2015}%
  \BibitemOpen
  \bibfield  {author} {\bibinfo {author} {\bibfnamefont {M.~A.}\ \bibnamefont
  {Laguna-Marco}}, \bibinfo {author} {\bibfnamefont {P.}~\bibnamefont
  {Kayser}}, \bibinfo {author} {\bibfnamefont {J.~A.}\ \bibnamefont {Alonso}},
  \bibinfo {author} {\bibfnamefont {M.~J.}\ \bibnamefont
  {Mart{\'{i}}nez-Lope}}, \bibinfo {author} {\bibfnamefont {M.}~\bibnamefont
  {van Veenendaal}}, \bibinfo {author} {\bibfnamefont {Y.}~\bibnamefont
  {Choi}},\ and\ \bibinfo {author} {\bibfnamefont {D.}~\bibnamefont {Haskel}},\
  }\bibfield  {title} {\bibinfo {title} {{Electronic structure, local
  magnetism, and spin-orbit effects of Ir(IV)-, Ir(V)-, and Ir(VI)-based
  compounds}},\ }\href {https://doi.org/10.1103/PhysRevB.91.214433} {\bibfield
  {journal} {\bibinfo  {journal} {Physical Review B}\ }\textbf {\bibinfo
  {volume} {91}},\ \bibinfo {pages} {214433} (\bibinfo {year}
  {2015})}\BibitemShut {NoStop}%
\bibitem [{\citenamefont {Gretarsson}\ \emph {et~al.}(2013)\citenamefont
  {Gretarsson}, \citenamefont {Clancy}, \citenamefont {Liu}, \citenamefont
  {Hill}, \citenamefont {Bozin}, \citenamefont {Singh}, \citenamefont {Manni},
  \citenamefont {Gegenwart}, \citenamefont {Kim}, \citenamefont {Said},
  \citenamefont {Casa}, \citenamefont {Gog}, \citenamefont {Upton},
  \citenamefont {Kim}, \citenamefont {Yu}, \citenamefont {Katukuri},
  \citenamefont {Hozoi}, \citenamefont {van~den Brink},\ and\ \citenamefont
  {Kim}}]{Gretarsson2013b}%
  \BibitemOpen
  \bibfield  {author} {\bibinfo {author} {\bibfnamefont {H.}~\bibnamefont
  {Gretarsson}}, \bibinfo {author} {\bibfnamefont {J.~P.}\ \bibnamefont
  {Clancy}}, \bibinfo {author} {\bibfnamefont {X.}~\bibnamefont {Liu}},
  \bibinfo {author} {\bibfnamefont {J.~P.}\ \bibnamefont {Hill}}, \bibinfo
  {author} {\bibfnamefont {E.}~\bibnamefont {Bozin}}, \bibinfo {author}
  {\bibfnamefont {Y.}~\bibnamefont {Singh}}, \bibinfo {author} {\bibfnamefont
  {S.}~\bibnamefont {Manni}}, \bibinfo {author} {\bibfnamefont
  {P.}~\bibnamefont {Gegenwart}}, \bibinfo {author} {\bibfnamefont
  {J.}~\bibnamefont {Kim}}, \bibinfo {author} {\bibfnamefont {A.~H.}\
  \bibnamefont {Said}}, \bibinfo {author} {\bibfnamefont {D.}~\bibnamefont
  {Casa}}, \bibinfo {author} {\bibfnamefont {T.}~\bibnamefont {Gog}}, \bibinfo
  {author} {\bibfnamefont {M.~H.}\ \bibnamefont {Upton}}, \bibinfo {author}
  {\bibfnamefont {H.-S.}\ \bibnamefont {Kim}}, \bibinfo {author} {\bibfnamefont
  {J.}~\bibnamefont {Yu}}, \bibinfo {author} {\bibfnamefont {V.~M.}\
  \bibnamefont {Katukuri}}, \bibinfo {author} {\bibfnamefont {L.}~\bibnamefont
  {Hozoi}}, \bibinfo {author} {\bibfnamefont {J.}~\bibnamefont {van~den
  Brink}},\ and\ \bibinfo {author} {\bibfnamefont {Y.-J.}\ \bibnamefont
  {Kim}},\ }\bibfield  {title} {\bibinfo {title} {{Crystal-Field Splitting and
  Correlation Effect on the Electronic Structure of \ch{A2IrO3}}},\ }\href
  {https://doi.org/10.1103/PhysRevLett.110.076402} {\bibfield  {journal}
  {\bibinfo  {journal} {Physical Review Letters}\ }\textbf {\bibinfo {volume}
  {110}},\ \bibinfo {pages} {076402} (\bibinfo {year} {2013})}\BibitemShut
  {NoStop}%
\bibitem [{\citenamefont {de~la Torre}\ \emph {et~al.}(2023)\citenamefont
  {de~la Torre}, \citenamefont {Zager}, \citenamefont {Bahrami}, \citenamefont
  {Upton}, \citenamefont {Kim}, \citenamefont {Fabbris}, \citenamefont {Lee},
  \citenamefont {Yang}, \citenamefont {Haskel}, \citenamefont {Tafti},\ and\
  \citenamefont {Plumb}}]{delaTorre2023}%
  \BibitemOpen
  \bibfield  {author} {\bibinfo {author} {\bibfnamefont {A.}~\bibnamefont
  {de~la Torre}}, \bibinfo {author} {\bibfnamefont {B.}~\bibnamefont {Zager}},
  \bibinfo {author} {\bibfnamefont {F.}~\bibnamefont {Bahrami}}, \bibinfo
  {author} {\bibfnamefont {M.~H.}\ \bibnamefont {Upton}}, \bibinfo {author}
  {\bibfnamefont {J.}~\bibnamefont {Kim}}, \bibinfo {author} {\bibfnamefont
  {G.}~\bibnamefont {Fabbris}}, \bibinfo {author} {\bibfnamefont {G.~H.}\
  \bibnamefont {Lee}}, \bibinfo {author} {\bibfnamefont {W.}~\bibnamefont
  {Yang}}, \bibinfo {author} {\bibfnamefont {D.}~\bibnamefont {Haskel}},
  \bibinfo {author} {\bibfnamefont {F.}~\bibnamefont {Tafti}},\ and\ \bibinfo
  {author} {\bibfnamefont {K.~W.}\ \bibnamefont {Plumb}},\ }\bibfield  {title}
  {\bibinfo {title} {Momentum-independent magnetic excitation continuum in the
  honeycomb iridate $\mathrm{H_3LiIr_2O_6}$},\ }\href
  {https://www.nature.com/articles/s41467-023-40769-x} {\bibfield  {journal}
  {\bibinfo  {journal} {Nature Communications}\ }\textbf {\bibinfo {volume}
  {14}},\ \bibinfo {pages} {5018} (\bibinfo {year} {2023})}\BibitemShut
  {NoStop}%
\bibitem [{\citenamefont {van Veenendaal}\ and\ \citenamefont
  {Haskel}(2022)}]{vanVeenendaal2022}%
  \BibitemOpen
  \bibfield  {author} {\bibinfo {author} {\bibfnamefont {M.}~\bibnamefont {van
  Veenendaal}}\ and\ \bibinfo {author} {\bibfnamefont {D.}~\bibnamefont
  {Haskel}},\ }\bibfield  {title} {\bibinfo {title} {Interpretation of ir
  \textit{L}-edge isotropic x-ray absorption spectra across the
  pressure-induced dimerization transition in hyperhoneycomb $\beta -
  \mathrm{Li_2IrO_3}$},\ }\href
  {https://journals.aps.org/prb/abstract/10.1103/PhysRevB.105.214420}
  {\bibfield  {journal} {\bibinfo  {journal} {Physical Review B}\ }\textbf
  {\bibinfo {volume} {105}},\ \bibinfo {pages} {214420} (\bibinfo {year}
  {2022})}\BibitemShut {NoStop}%
\bibitem [{\citenamefont {Mazin}\ \emph {et~al.}(2012)\citenamefont {Mazin},
  \citenamefont {Jeschke}, \citenamefont {Foyevtsova}, \citenamefont
  {Valent{\'{i}}},\ and\ \citenamefont {Khomskii}}]{Mazin2012}%
  \BibitemOpen
  \bibfield  {author} {\bibinfo {author} {\bibfnamefont {I.~I.}\ \bibnamefont
  {Mazin}}, \bibinfo {author} {\bibfnamefont {H.~O.}\ \bibnamefont {Jeschke}},
  \bibinfo {author} {\bibfnamefont {K.}~\bibnamefont {Foyevtsova}}, \bibinfo
  {author} {\bibfnamefont {R.}~\bibnamefont {Valent{\'{i}}}},\ and\ \bibinfo
  {author} {\bibfnamefont {D.~I.}\ \bibnamefont {Khomskii}},\ }\bibfield
  {title} {\bibinfo {title} {{\ch{Na2IrO3} as a Molecular Orbital Crystal}},\
  }\href {https://doi.org/10.1103/PhysRevLett.109.197201} {\bibfield  {journal}
  {\bibinfo  {journal} {Physical Review Letters}\ }\textbf {\bibinfo {volume}
  {109}},\ \bibinfo {pages} {197201} (\bibinfo {year} {2012})}\BibitemShut
  {NoStop}%
\bibitem [{\citenamefont {Revelli}\ \emph {et~al.}(2019)\citenamefont
  {Revelli}, \citenamefont {Moretti~Sala}, \citenamefont {Monaco},
  \citenamefont {Becker}, \citenamefont {Bohaty}, \citenamefont {Hermanns},
  \citenamefont {Koethe}, \citenamefont {Frohlich}, \citenamefont
  {Warzanowski}, \citenamefont {Lorenz}, \citenamefont {Streltsov},
  \citenamefont {van Loosdrecht}, \citenamefont {Khomskii}, \citenamefont
  {van~den Brink},\ and\ \citenamefont {Gruninger}}]{Revelli2019}%
  \BibitemOpen
  \bibfield  {author} {\bibinfo {author} {\bibfnamefont {A.}~\bibnamefont
  {Revelli}}, \bibinfo {author} {\bibfnamefont {M.}~\bibnamefont
  {Moretti~Sala}}, \bibinfo {author} {\bibfnamefont {G.}~\bibnamefont
  {Monaco}}, \bibinfo {author} {\bibfnamefont {P.}~\bibnamefont {Becker}},
  \bibinfo {author} {\bibfnamefont {L.}~\bibnamefont {Bohaty}}, \bibinfo
  {author} {\bibfnamefont {M.}~\bibnamefont {Hermanns}}, \bibinfo {author}
  {\bibfnamefont {T.~C.}\ \bibnamefont {Koethe}}, \bibinfo {author}
  {\bibfnamefont {T.}~\bibnamefont {Frohlich}}, \bibinfo {author}
  {\bibfnamefont {P.}~\bibnamefont {Warzanowski}}, \bibinfo {author}
  {\bibfnamefont {T.}~\bibnamefont {Lorenz}}, \bibinfo {author} {\bibfnamefont
  {S.~V.}\ \bibnamefont {Streltsov}}, \bibinfo {author} {\bibfnamefont
  {P.~H.~M.}\ \bibnamefont {van Loosdrecht}}, \bibinfo {author} {\bibfnamefont
  {D.~I.}\ \bibnamefont {Khomskii}}, \bibinfo {author} {\bibfnamefont
  {J.}~\bibnamefont {van~den Brink}},\ and\ \bibinfo {author} {\bibfnamefont
  {M.}~\bibnamefont {Gruninger}},\ }\bibfield  {title} {\bibinfo {title}
  {Resonant inelastic x-ray incarnation of young's double-slit experiment},\
  }\href {https://doi.org/10.1126/sciadv.aav4020} {\bibfield  {journal}
  {\bibinfo  {journal} {Science Advances}\ }\textbf {\bibinfo {volume} {5}},\
  \bibinfo {pages} {eaav4020} (\bibinfo {year} {2019})}\BibitemShut {NoStop}%
\bibitem [{\citenamefont {Wang}\ \emph {et~al.}(2019)\citenamefont {Wang},
  \citenamefont {Wang}, \citenamefont {Kim}, \citenamefont {Upton},
  \citenamefont {Casa}, \citenamefont {Gog}, \citenamefont {Cao}, \citenamefont
  {Kotliar}, \citenamefont {Dean},\ and\ \citenamefont {Liu}}]{Wang2019b}%
  \BibitemOpen
  \bibfield  {author} {\bibinfo {author} {\bibfnamefont {Y.}~\bibnamefont
  {Wang}}, \bibinfo {author} {\bibfnamefont {R.}~\bibnamefont {Wang}}, \bibinfo
  {author} {\bibfnamefont {J.}~\bibnamefont {Kim}}, \bibinfo {author}
  {\bibfnamefont {M.~H.}\ \bibnamefont {Upton}}, \bibinfo {author}
  {\bibfnamefont {D.}~\bibnamefont {Casa}}, \bibinfo {author} {\bibfnamefont
  {T.}~\bibnamefont {Gog}}, \bibinfo {author} {\bibfnamefont {G.}~\bibnamefont
  {Cao}}, \bibinfo {author} {\bibfnamefont {G.}~\bibnamefont {Kotliar}},
  \bibinfo {author} {\bibfnamefont {M.~P.~M.}\ \bibnamefont {Dean}},\ and\
  \bibinfo {author} {\bibfnamefont {X.}~\bibnamefont {Liu}},\ }\bibfield
  {title} {\bibinfo {title} {Direct detection of dimer orbitals in
  \ch{Ba5AlIr2O11}},\ }\href {https://doi.org/10.1103/PhysRevLett.122.106401}
  {\bibfield  {journal} {\bibinfo  {journal} {Physical Review Letters}\
  }\textbf {\bibinfo {volume} {122}},\ \bibinfo {pages} {106401} (\bibinfo
  {year} {2019})}\BibitemShut {NoStop}%
\bibitem [{\citenamefont {Mackenzie}(2017)}]{Mackenzie2017}%
  \BibitemOpen
  \bibfield  {author} {\bibinfo {author} {\bibfnamefont {A.~P.}\ \bibnamefont
  {Mackenzie}},\ }\bibfield  {title} {\bibinfo {title} {The properties of
  ultrapure delafossite metals},\ }\href
  {https://doi.org/10.1088/1361-6633/aa50e5} {\bibfield  {journal} {\bibinfo
  {journal} {Reports on Progress on Physics}\ }\textbf {\bibinfo {volume}
  {80}},\ \bibinfo {pages} {032501} (\bibinfo {year} {2017})}\BibitemShut
  {NoStop}%
\bibitem [{\citenamefont {Xu}\ \emph {et~al.}(2016)\citenamefont {Xu},
  \citenamefont {Hearne},\ and\ \citenamefont {Pasternak}}]{Xu2010}%
  \BibitemOpen
  \bibfield  {author} {\bibinfo {author} {\bibfnamefont {W.~M.}\ \bibnamefont
  {Xu}}, \bibinfo {author} {\bibfnamefont {G.~R.}\ \bibnamefont {Hearne}},\
  and\ \bibinfo {author} {\bibfnamefont {M.~P.}\ \bibnamefont {Pasternak}},\
  }\bibfield  {title} {\bibinfo {title} {$\mathrm{CuFeO_2}$ at a megabar:
  Stabilization of a mixed-valence low-spin magnetic semiconducting ground
  state},\ }\href {https://doi.org/10.1103/PhysRevB.94.035155} {\bibfield
  {journal} {\bibinfo  {journal} {Physical Review B}\ }\textbf {\bibinfo
  {volume} {94}},\ \bibinfo {pages} {035155} (\bibinfo {year}
  {2016})}\BibitemShut {NoStop}%
\bibitem [{\citenamefont {Xu}\ \emph {et~al.}(2010)\citenamefont {Xu},
  \citenamefont {Rozenberg}, \citenamefont {Pasternak}, \citenamefont
  {Kertzer}, \citenamefont {Kurnosov}, \citenamefont {Dubrovinsky},
  \citenamefont {Pascarelli}, \citenamefont {Munoz}, \citenamefont {Vaccari},
  \citenamefont {Hanfland},\ and\ \citenamefont {Jeanloz}}]{Xu2016}%
  \BibitemOpen
  \bibfield  {author} {\bibinfo {author} {\bibfnamefont {W.~M.}\ \bibnamefont
  {Xu}}, \bibinfo {author} {\bibfnamefont {G.~K.}\ \bibnamefont {Rozenberg}},
  \bibinfo {author} {\bibfnamefont {M.~P.}\ \bibnamefont {Pasternak}}, \bibinfo
  {author} {\bibfnamefont {M.}~\bibnamefont {Kertzer}}, \bibinfo {author}
  {\bibfnamefont {A.}~\bibnamefont {Kurnosov}}, \bibinfo {author}
  {\bibfnamefont {L.~S.}\ \bibnamefont {Dubrovinsky}}, \bibinfo {author}
  {\bibfnamefont {S.}~\bibnamefont {Pascarelli}}, \bibinfo {author}
  {\bibfnamefont {M.}~\bibnamefont {Munoz}}, \bibinfo {author} {\bibfnamefont
  {M.}~\bibnamefont {Vaccari}}, \bibinfo {author} {\bibfnamefont
  {M.}~\bibnamefont {Hanfland}},\ and\ \bibinfo {author} {\bibfnamefont
  {R.}~\bibnamefont {Jeanloz}},\ }\bibfield  {title} {\bibinfo {title}
  {Pressure-induced $\mathrm{Fe \leftrightarrow Cu}$ cationic valence exchange
  and its structural consequences: High-pressure studies of delafossite
  $\mathrm{CuFeO_2}$},\ }\href {https://doi.org/10.1103/PhysRevB.81.104110}
  {\bibfield  {journal} {\bibinfo  {journal} {Physical Review B}\ }\textbf
  {\bibinfo {volume} {81}},\ \bibinfo {pages} {104110} (\bibinfo {year}
  {2010})}\BibitemShut {NoStop}%
\bibitem [{\citenamefont {Garg}\ and\ \citenamefont {Rao}(2018)}]{Garg2018}%
  \BibitemOpen
  \bibfield  {author} {\bibinfo {author} {\bibfnamefont {A.}~\bibnamefont
  {Garg}}\ and\ \bibinfo {author} {\bibfnamefont {R.}~\bibnamefont {Rao}},\
  }\bibfield  {title} {\bibinfo {title} {Copper delafossites under high
  pressure — a brief review of xrd and raman spectroscopic studies},\ }\href
  {https://doi.org/10.3390/cryst8060255} {\bibfield  {journal} {\bibinfo
  {journal} {Crystals}\ }\textbf {\bibinfo {volume} {8}},\ \bibinfo {pages}
  {255} (\bibinfo {year} {2018})}\BibitemShut {NoStop}%
\bibitem [{\citenamefont {Levy}\ \emph {et~al.}(2020)\citenamefont {Levy},
  \citenamefont {Greenberg}, \citenamefont {Layek}, \citenamefont {Pasternak},
  \citenamefont {Kantor}, \citenamefont {Pascarelli}, \citenamefont {Marini},
  \citenamefont {Konopkova},\ and\ \citenamefont {Rozenberg}}]{Levy2020}%
  \BibitemOpen
  \bibfield  {author} {\bibinfo {author} {\bibfnamefont {D.}~\bibnamefont
  {Levy}}, \bibinfo {author} {\bibfnamefont {E.}~\bibnamefont {Greenberg}},
  \bibinfo {author} {\bibfnamefont {S.}~\bibnamefont {Layek}}, \bibinfo
  {author} {\bibfnamefont {M.~P.}\ \bibnamefont {Pasternak}}, \bibinfo {author}
  {\bibfnamefont {I.}~\bibnamefont {Kantor}}, \bibinfo {author} {\bibfnamefont
  {S.}~\bibnamefont {Pascarelli}}, \bibinfo {author} {\bibfnamefont
  {C.}~\bibnamefont {Marini}}, \bibinfo {author} {\bibfnamefont
  {Z.}~\bibnamefont {Konopkova}},\ and\ \bibinfo {author} {\bibfnamefont
  {G.~K.}\ \bibnamefont {Rozenberg}},\ }\bibfield  {title} {\bibinfo {title}
  {High-pressure structural and electronic properties of $\mathrm{CuMO_2}$
  ($\mathrm{M=Cr, Mn}$) delafossite-type oxides},\ }\href
  {https://doi.org/10.1103/PhysRevB.101.245121} {\bibfield  {journal} {\bibinfo
   {journal} {Physical Review B}\ }\textbf {\bibinfo {volume} {101}},\ \bibinfo
  {pages} {245121} (\bibinfo {year} {2020})}\BibitemShut {NoStop}%
\bibitem [{\citenamefont {Lawler}\ \emph {et~al.}(2021)\citenamefont {Lawler},
  \citenamefont {Smith}, \citenamefont {Evans}, \citenamefont {Dos~Santos},
  \citenamefont {Molaison}, \citenamefont {Bos}, \citenamefont {Mutka},
  \citenamefont {Henry}, \citenamefont {Argyriou}, \citenamefont {Salamat},\
  and\ \citenamefont {Kimber}}]{Lawler2021}%
  \BibitemOpen
  \bibfield  {author} {\bibinfo {author} {\bibfnamefont {K.~V.}\ \bibnamefont
  {Lawler}}, \bibinfo {author} {\bibfnamefont {D.}~\bibnamefont {Smith}},
  \bibinfo {author} {\bibfnamefont {S.~R.}\ \bibnamefont {Evans}}, \bibinfo
  {author} {\bibfnamefont {A.~M.}\ \bibnamefont {Dos~Santos}}, \bibinfo
  {author} {\bibfnamefont {J.~J.}\ \bibnamefont {Molaison}}, \bibinfo {author}
  {\bibfnamefont {J.~G.}\ \bibnamefont {Bos}}, \bibinfo {author} {\bibfnamefont
  {H.}~\bibnamefont {Mutka}}, \bibinfo {author} {\bibfnamefont {P.~F.}\
  \bibnamefont {Henry}}, \bibinfo {author} {\bibfnamefont {D.~N.}\ \bibnamefont
  {Argyriou}}, \bibinfo {author} {\bibfnamefont {A.}~\bibnamefont {Salamat}},\
  and\ \bibinfo {author} {\bibfnamefont {S.~A.~J.}\ \bibnamefont {Kimber}},\
  }\bibfield  {title} {\bibinfo {title} {Decoupling lattice and magnetic
  instabilities in frustrated $\mathrm{CuMnO_2}$},\ }\href
  {https://doi.org/10.1021/acs.inorgchem.1c00435} {\bibfield  {journal}
  {\bibinfo  {journal} {Inorganic Chemistry}\ }\textbf {\bibinfo {volume}
  {60}},\ \bibinfo {pages} {6004} (\bibinfo {year} {2021})}\BibitemShut
  {NoStop}%
\bibitem [{\citenamefont {Haraguchi}\ \emph {et~al.}(2024)\citenamefont
  {Haraguchi}, \citenamefont {Nishio-Hamane}, \citenamefont {Matsuo},
  \citenamefont {Kindo},\ and\ \citenamefont {Katori}}]{Haraguchi2024}%
  \BibitemOpen
  \bibfield  {author} {\bibinfo {author} {\bibfnamefont {Y.}~\bibnamefont
  {Haraguchi}}, \bibinfo {author} {\bibfnamefont {D.}~\bibnamefont
  {Nishio-Hamane}}, \bibinfo {author} {\bibfnamefont {A.}~\bibnamefont
  {Matsuo}}, \bibinfo {author} {\bibfnamefont {K.}~\bibnamefont {Kindo}},\ and\
  \bibinfo {author} {\bibfnamefont {H.~A.}\ \bibnamefont {Katori}},\ }\bibfield
   {title} {\bibinfo {title} {High-temperature magnetic anomaly via suppression
  of antisite disorder through synthesis route modification in a kitaev
  candidate \ch{Cu2IrO3}},\ }\href {https://doi.org/10.1088/1361-648X/ad5d3a}
  {\bibfield  {journal} {\bibinfo  {journal} {Journal of Physics: Condensed
  Matter}\ }\textbf {\bibinfo {volume} {36}},\ \bibinfo {pages} {405801}
  (\bibinfo {year} {2024})}\BibitemShut {NoStop}%
\bibitem [{\citenamefont {Abramchuk}\ \emph {et~al.}(2018)\citenamefont
  {Abramchuk}, \citenamefont {Lebedev}, \citenamefont {Hellman}, \citenamefont
  {Bahrami}, \citenamefont {Mordvinova}, \citenamefont {Krizan}, \citenamefont
  {Metz}, \citenamefont {Broido},\ and\ \citenamefont {Tafti}}]{Abramchuk2018}%
  \BibitemOpen
  \bibfield  {author} {\bibinfo {author} {\bibfnamefont {M.}~\bibnamefont
  {Abramchuk}}, \bibinfo {author} {\bibfnamefont {O.~I.}\ \bibnamefont
  {Lebedev}}, \bibinfo {author} {\bibfnamefont {O.}~\bibnamefont {Hellman}},
  \bibinfo {author} {\bibfnamefont {F.}~\bibnamefont {Bahrami}}, \bibinfo
  {author} {\bibfnamefont {N.~E.}\ \bibnamefont {Mordvinova}}, \bibinfo
  {author} {\bibfnamefont {J.~W.}\ \bibnamefont {Krizan}}, \bibinfo {author}
  {\bibfnamefont {K.~R.}\ \bibnamefont {Metz}}, \bibinfo {author}
  {\bibfnamefont {D.}~\bibnamefont {Broido}},\ and\ \bibinfo {author}
  {\bibfnamefont {F.}~\bibnamefont {Tafti}},\ }\bibfield  {title} {\bibinfo
  {title} {Crystal chemistry and phonon heat capacity in quaternary honeycomb
  delafossites: $\mathrm{Cu[Li_{1/3}Sn_{2/3}]O_2}$ and
  $\mathrm{Cu[Na_{1/3}Sn_{2/3}]O_2}$},\ }\href
  {https://doi.org/10.1021/acs.inorgchem.8b01866} {\bibfield  {journal}
  {\bibinfo  {journal} {Inorganic Chemistry}\ }\textbf {\bibinfo {volume}
  {57}},\ \bibinfo {pages} {12709} (\bibinfo {year} {2018})}\BibitemShut
  {NoStop}%
\bibitem [{\citenamefont {Fisch}\ \emph {et~al.}(2015)\citenamefont {Fisch},
  \citenamefont {Lanza}, \citenamefont {Boldyreva}, \citenamefont {Macchi},\
  and\ \citenamefont {Casati}}]{Fisch2015}%
  \BibitemOpen
  \bibfield  {author} {\bibinfo {author} {\bibfnamefont {M.}~\bibnamefont
  {Fisch}}, \bibinfo {author} {\bibfnamefont {A.}~\bibnamefont {Lanza}},
  \bibinfo {author} {\bibfnamefont {E.}~\bibnamefont {Boldyreva}}, \bibinfo
  {author} {\bibfnamefont {P.}~\bibnamefont {Macchi}},\ and\ \bibinfo {author}
  {\bibfnamefont {N.}~\bibnamefont {Casati}},\ }\bibfield  {title} {\bibinfo
  {title} {Kinetic control of high-pressure solid-state phase transitions: A
  case study on l-serine},\ }\href {https://doi.org/10.1021/acs.jpcc.5b05838}
  {\bibfield  {journal} {\bibinfo  {journal} {The Journal of Physical Chemistry
  C}\ }\textbf {\bibinfo {volume} {119}},\ \bibinfo {pages} {18611} (\bibinfo
  {year} {2015})}\BibitemShut {NoStop}%
\bibitem [{\citenamefont {Haskel}\ \emph {et~al.}(2008)\citenamefont {Haskel},
  \citenamefont {Tseng}, \citenamefont {Souza-Neto}, \citenamefont {Lang},
  \citenamefont {Sinogeikin}, \citenamefont {Mudryk}, \citenamefont
  {Gschneidner},\ and\ \citenamefont {Pecharsky}}]{Haskel2008}%
  \BibitemOpen
  \bibfield  {author} {\bibinfo {author} {\bibfnamefont {D.}~\bibnamefont
  {Haskel}}, \bibinfo {author} {\bibfnamefont {Y.~C.}\ \bibnamefont {Tseng}},
  \bibinfo {author} {\bibfnamefont {N.~M.}\ \bibnamefont {Souza-Neto}},
  \bibinfo {author} {\bibfnamefont {J.~C.}\ \bibnamefont {Lang}}, \bibinfo
  {author} {\bibfnamefont {S.}~\bibnamefont {Sinogeikin}}, \bibinfo {author}
  {\bibfnamefont {Y.}~\bibnamefont {Mudryk}}, \bibinfo {author} {\bibfnamefont
  {K.~A.}\ \bibnamefont {Gschneidner}},\ and\ \bibinfo {author} {\bibfnamefont
  {V.~K.}\ \bibnamefont {Pecharsky}},\ }\bibfield  {title} {\bibinfo {title}
  {{Magnetic spectroscopy at high pressures using X-ray magnetic circular
  dichroism}},\ }\href {https://doi.org/10.1080/08957950802020307} {\bibfield
  {journal} {\bibinfo  {journal} {High Pressure Research}\ }\textbf {\bibinfo
  {volume} {28}},\ \bibinfo {pages} {185} (\bibinfo {year} {2008})}\BibitemShut
  {NoStop}%
\bibitem [{\citenamefont {Dewaele}\ \emph {et~al.}(2008)\citenamefont
  {Dewaele}, \citenamefont {Torrent}, \citenamefont {Loubeyre},\ and\
  \citenamefont {Mezouar}}]{Dewaele2008}%
  \BibitemOpen
  \bibfield  {author} {\bibinfo {author} {\bibfnamefont {A.}~\bibnamefont
  {Dewaele}}, \bibinfo {author} {\bibfnamefont {M.}~\bibnamefont {Torrent}},
  \bibinfo {author} {\bibfnamefont {P.}~\bibnamefont {Loubeyre}},\ and\
  \bibinfo {author} {\bibfnamefont {M.}~\bibnamefont {Mezouar}},\ }\bibfield
  {title} {\bibinfo {title} {{Compression curves of transition metals in the
  Mbar range: Experiments and projector augmented-wave calculations}},\ }\href
  {https://doi.org/10.1103/PhysRevB.78.104102} {\bibfield  {journal} {\bibinfo
  {journal} {Physical Review B}\ }\textbf {\bibinfo {volume} {78}},\ \bibinfo
  {pages} {104102} (\bibinfo {year} {2008})}\BibitemShut {NoStop}%
\bibitem [{\citenamefont {Rivers}\ \emph {et~al.}(2008)\citenamefont {Rivers},
  \citenamefont {Prakapenka}, \citenamefont {Kubo}, \citenamefont {Pullins},
  \citenamefont {Holl},\ and\ \citenamefont {Jacobsen}}]{Rivers2008}%
  \BibitemOpen
  \bibfield  {author} {\bibinfo {author} {\bibfnamefont {M.}~\bibnamefont
  {Rivers}}, \bibinfo {author} {\bibfnamefont {V.}~\bibnamefont {Prakapenka}},
  \bibinfo {author} {\bibfnamefont {A.}~\bibnamefont {Kubo}}, \bibinfo {author}
  {\bibfnamefont {C.}~\bibnamefont {Pullins}}, \bibinfo {author} {\bibfnamefont
  {C.}~\bibnamefont {Holl}},\ and\ \bibinfo {author} {\bibfnamefont
  {S.}~\bibnamefont {Jacobsen}},\ }\bibfield  {title} {\bibinfo {title} {{The
  COMPRES/GSECARS gas-loading system for diamond anvil cells at the Advanced
  Photon Source}},\ }\href {https://doi.org/10.1080/08957950802333593}
  {\bibfield  {journal} {\bibinfo  {journal} {High Pressure Research}\ }\textbf
  {\bibinfo {volume} {28}},\ \bibinfo {pages} {273} (\bibinfo {year}
  {2008})}\BibitemShut {NoStop}%
\bibitem [{\citenamefont {Newville}(2013)}]{Newville2013}%
  \BibitemOpen
  \bibfield  {author} {\bibinfo {author} {\bibfnamefont {M.}~\bibnamefont
  {Newville}},\ }\bibfield  {title} {\bibinfo {title} {{Larch: An Analysis
  Package for XAFS and Related Spectroscopies}},\ }\href
  {https://doi.org/10.1088/1742-6596/430/1/012007} {\bibfield  {journal}
  {\bibinfo  {journal} {Journal of Physics: Conference Series}\ }\textbf
  {\bibinfo {volume} {430}},\ \bibinfo {pages} {012007} (\bibinfo {year}
  {2013})}\BibitemShut {NoStop}%
\bibitem [{\citenamefont {Kim}(2016)}]{Kim2016b}%
  \BibitemOpen
  \bibfield  {author} {\bibinfo {author} {\bibfnamefont {J.}~\bibnamefont
  {Kim}},\ }\bibfield  {title} {\bibinfo {title} {{Advances in high-resolution
  RIXS for the study of excitation spectra under high pressure}},\ }\href
  {https://doi.org/10.1080/08957959.2016.1212990} {\bibfield  {journal}
  {\bibinfo  {journal} {High Pressure Research}\ }\textbf {\bibinfo {volume}
  {36}},\ \bibinfo {pages} {391} (\bibinfo {year} {2016})}\BibitemShut
  {NoStop}%
\bibitem [{\citenamefont {Holzapfel}\ \emph {et~al.}(2001)\citenamefont
  {Holzapfel}, \citenamefont {Hartwig},\ and\ \citenamefont
  {Sievers}}]{Holzapfel2001}%
  \BibitemOpen
  \bibfield  {author} {\bibinfo {author} {\bibfnamefont {W.~B.}\ \bibnamefont
  {Holzapfel}}, \bibinfo {author} {\bibfnamefont {M.}~\bibnamefont {Hartwig}},\
  and\ \bibinfo {author} {\bibfnamefont {W.}~\bibnamefont {Sievers}},\
  }\bibfield  {title} {\bibinfo {title} {{Equations of state for Cu, Ag, and Au
  for wide ranges in temperature and pressure up to 500 GPa and above}},\
  }\href {https://doi.org/10.1063/1.1370170} {\bibfield  {journal} {\bibinfo
  {journal} {Journal of Physical and Chemical Reference Data}\ }\textbf
  {\bibinfo {volume} {30}},\ \bibinfo {pages} {515} (\bibinfo {year}
  {2001})}\BibitemShut {NoStop}%
\bibitem [{\citenamefont {Ragan}\ \emph {et~al.}(1992)\citenamefont {Ragan},
  \citenamefont {Gustavsen},\ and\ \citenamefont {Schiferl}}]{Ragan1992}%
  \BibitemOpen
  \bibfield  {author} {\bibinfo {author} {\bibfnamefont {D.~D.}\ \bibnamefont
  {Ragan}}, \bibinfo {author} {\bibfnamefont {R.}~\bibnamefont {Gustavsen}},\
  and\ \bibinfo {author} {\bibfnamefont {D.}~\bibnamefont {Schiferl}},\
  }\bibfield  {title} {\bibinfo {title} {{Calibration of the ruby $R_1$ and
  $R_2$ fluorescence shifts as a function of temperature from 0 to 600 K}},\
  }\href {https://doi.org/10.1063/1.351951} {\bibfield  {journal} {\bibinfo
  {journal} {Journal of Applied Physics}\ }\textbf {\bibinfo {volume} {72}},\
  \bibinfo {pages} {5539} (\bibinfo {year} {1992})}\BibitemShut {NoStop}%
\bibitem [{\citenamefont {Prescher}\ and\ \citenamefont
  {Prakapenka}(2015)}]{Prescher2015}%
  \BibitemOpen
  \bibfield  {author} {\bibinfo {author} {\bibfnamefont {C.}~\bibnamefont
  {Prescher}}\ and\ \bibinfo {author} {\bibfnamefont {V.~B.}\ \bibnamefont
  {Prakapenka}},\ }\bibfield  {title} {\bibinfo {title} {{DIOPTAS : a program
  for reduction of two-dimensional X-ray diffraction data and data
  exploration}},\ }\href {https://doi.org/10.1080/08957959.2015.1059835}
  {\bibfield  {journal} {\bibinfo  {journal} {High Pressure Research}\ }\textbf
  {\bibinfo {volume} {35}},\ \bibinfo {pages} {223} (\bibinfo {year}
  {2015})}\BibitemShut {NoStop}%
\bibitem [{\citenamefont {Chijioke}\ \emph {et~al.}(2005)\citenamefont
  {Chijioke}, \citenamefont {Nellis}, \citenamefont {Soldatov},\ and\
  \citenamefont {Silvera}}]{chijioke_ruby_2005}%
  \BibitemOpen
  \bibfield  {author} {\bibinfo {author} {\bibfnamefont {A.~D.}\ \bibnamefont
  {Chijioke}}, \bibinfo {author} {\bibfnamefont {W.~J.}\ \bibnamefont
  {Nellis}}, \bibinfo {author} {\bibfnamefont {A.}~\bibnamefont {Soldatov}},\
  and\ \bibinfo {author} {\bibfnamefont {I.~F.}\ \bibnamefont {Silvera}},\
  }\bibfield  {title} {\bibinfo {title} {The ruby pressure standard to {150
  GPa}},\ }\href {https://doi.org/10.1063/1.2135877} {\bibfield  {journal}
  {\bibinfo  {journal} {Journal of Applied Physics}\ }\textbf {\bibinfo
  {volume} {98}},\ \bibinfo {pages} {114905} (\bibinfo {year}
  {2005})}\BibitemShut {NoStop}%
\bibitem [{\citenamefont {Weir}\ \emph {et~al.}(2000)\citenamefont {Weir},
  \citenamefont {Akella}, \citenamefont {Aracne-Ruddle}, \citenamefont
  {Vohra},\ and\ \citenamefont {Catledge}}]{Weir_DesignerAnvil_2000}%
  \BibitemOpen
  \bibfield  {author} {\bibinfo {author} {\bibfnamefont {S.~T.}\ \bibnamefont
  {Weir}}, \bibinfo {author} {\bibfnamefont {J.}~\bibnamefont {Akella}},
  \bibinfo {author} {\bibfnamefont {C.}~\bibnamefont {Aracne-Ruddle}}, \bibinfo
  {author} {\bibfnamefont {Y.~K.}\ \bibnamefont {Vohra}},\ and\ \bibinfo
  {author} {\bibfnamefont {S.~A.}\ \bibnamefont {Catledge}},\ }\bibfield
  {title} {\bibinfo {title} {Epitaxial diamond encapsulation of metal
  microprobes for high pressure experiments},\ }\href
  {https://doi.org/10.1063/1.1326838} {\bibfield  {journal} {\bibinfo
  {journal} {Applied Physics Letters}\ }\textbf {\bibinfo {volume} {77}},\
  \bibinfo {pages} {3400} (\bibinfo {year} {2000})}\BibitemShut {NoStop}%
\bibitem [{\citenamefont {Lim}\ \emph {et~al.}(2021)\citenamefont {Lim},
  \citenamefont {Hire}, \citenamefont {Quan}, \citenamefont {Kim},
  \citenamefont {Fanfarillo}, \citenamefont {Xie}, \citenamefont {Kumar},
  \citenamefont {Park}, \citenamefont {Hemley}, \citenamefont {Vohra},
  \citenamefont {Hennig}, \citenamefont {Hirschfeld}, \citenamefont {Stewart},\
  and\ \citenamefont {Hamlin}}]{Lim2021}%
  \BibitemOpen
  \bibfield  {author} {\bibinfo {author} {\bibfnamefont {J.}~\bibnamefont
  {Lim}}, \bibinfo {author} {\bibfnamefont {A.~C.}\ \bibnamefont {Hire}},
  \bibinfo {author} {\bibfnamefont {Y.}~\bibnamefont {Quan}}, \bibinfo {author}
  {\bibfnamefont {J.}~\bibnamefont {Kim}}, \bibinfo {author} {\bibfnamefont
  {L.}~\bibnamefont {Fanfarillo}}, \bibinfo {author} {\bibfnamefont {S.~R.}\
  \bibnamefont {Xie}}, \bibinfo {author} {\bibfnamefont {R.~S.}\ \bibnamefont
  {Kumar}}, \bibinfo {author} {\bibfnamefont {C.}~\bibnamefont {Park}},
  \bibinfo {author} {\bibfnamefont {R.~J.}\ \bibnamefont {Hemley}}, \bibinfo
  {author} {\bibfnamefont {Y.~K.}\ \bibnamefont {Vohra}}, \bibinfo {author}
  {\bibfnamefont {R.~G.}\ \bibnamefont {Hennig}}, \bibinfo {author}
  {\bibfnamefont {P.~J.}\ \bibnamefont {Hirschfeld}}, \bibinfo {author}
  {\bibfnamefont {G.~R.}\ \bibnamefont {Stewart}},\ and\ \bibinfo {author}
  {\bibfnamefont {J.~J.}\ \bibnamefont {Hamlin}},\ }\bibfield  {title}
  {\bibinfo {title} {High-pressure study of the low-\uppercase{Z} rich
  superconductor $\mathrm{Be_{22}Re}$},\ }\href
  {https://doi.org/10.1103/physrevb.104.064505} {\bibfield  {journal} {\bibinfo
   {journal} {Physical Review B}\ }\textbf {\bibinfo {volume} {104}},\ \bibinfo
  {pages} {064505} (\bibinfo {year} {2021})}\BibitemShut {NoStop}%
\bibitem [{\citenamefont {{van der Laan}}\ and\ \citenamefont
  {Figueroa}(2014)}]{vanderLaan2014}%
  \BibitemOpen
  \bibfield  {author} {\bibinfo {author} {\bibfnamefont {G.}~\bibnamefont {{van
  der Laan}}}\ and\ \bibinfo {author} {\bibfnamefont {A.~I.}\ \bibnamefont
  {Figueroa}},\ }\bibfield  {title} {\bibinfo {title} {X-ray magnetic circular
  dichroism—a versatile tool to study magnetism},\ }\href
  {https://doi.org/10.1016/j.ccr.2014.03.018} {\bibfield  {journal} {\bibinfo
  {journal} {Coordination Chemistry Reviews}\ }\textbf {\bibinfo {volume}
  {277-278}},\ \bibinfo {pages} {95} (\bibinfo {year} {2014})}\BibitemShut
  {NoStop}%
\bibitem [{\citenamefont {Nagamatsu}\ \emph {et~al.}(2004)\citenamefont
  {Nagamatsu}, \citenamefont {Matsumoto}, \citenamefont {Fujikawa},
  \citenamefont {Ishiji},\ and\ \citenamefont {Hashizume}}]{Nagamatsu2004}%
  \BibitemOpen
  \bibfield  {author} {\bibinfo {author} {\bibfnamefont {S.}~\bibnamefont
  {Nagamatsu}}, \bibinfo {author} {\bibfnamefont {H.}~\bibnamefont
  {Matsumoto}}, \bibinfo {author} {\bibfnamefont {T.}~\bibnamefont {Fujikawa}},
  \bibinfo {author} {\bibfnamefont {K.}~\bibnamefont {Ishiji}},\ and\ \bibinfo
  {author} {\bibfnamefont {H.}~\bibnamefont {Hashizume}},\ }\bibfield  {title}
  {\bibinfo {title} {Measurement and multiple-scattering calculation of cu
  \textit{K}-edge x-ray magnetic circular dichroism spectra from an
  exchange-coupled $\mathrm{Co}/\mathrm{Cu}$ multilayer},\ }\href
  {https://doi.org/10.1103/PhysRevB.70.174442} {\bibfield  {journal} {\bibinfo
  {journal} {Physical Review B}\ }\textbf {\bibinfo {volume} {70}},\ \bibinfo
  {pages} {174442} (\bibinfo {year} {2004})}\BibitemShut {NoStop}%
\bibitem [{\citenamefont {Ivanov}\ \emph {et~al.}(2018)\citenamefont {Ivanov},
  \citenamefont {Ivanov}, \citenamefont {Menushenkov}, \citenamefont {Wilhelm},
  \citenamefont {Rogalev}, \citenamefont {Puri}, \citenamefont {Joseph},
  \citenamefont {Xu}, \citenamefont {Marcelli},\ and\ \citenamefont
  {Bianconi}}]{Ivanov2018}%
  \BibitemOpen
  \bibfield  {author} {\bibinfo {author} {\bibfnamefont {A.~A.}\ \bibnamefont
  {Ivanov}}, \bibinfo {author} {\bibfnamefont {V.~G.}\ \bibnamefont {Ivanov}},
  \bibinfo {author} {\bibfnamefont {A.~P.}\ \bibnamefont {Menushenkov}},
  \bibinfo {author} {\bibfnamefont {F.}~\bibnamefont {Wilhelm}}, \bibinfo
  {author} {\bibfnamefont {A.}~\bibnamefont {Rogalev}}, \bibinfo {author}
  {\bibfnamefont {A.}~\bibnamefont {Puri}}, \bibinfo {author} {\bibfnamefont
  {B.}~\bibnamefont {Joseph}}, \bibinfo {author} {\bibfnamefont
  {W.}~\bibnamefont {Xu}}, \bibinfo {author} {\bibfnamefont {A.}~\bibnamefont
  {Marcelli}},\ and\ \bibinfo {author} {\bibfnamefont {A.}~\bibnamefont
  {Bianconi}},\ }\bibfield  {title} {\bibinfo {title} {Local noncentrosymmetric
  structure of $\mathrm{Bi_2Sr_2CaCu_2O_{8+y}}$ by x-ray magnetic circular
  dichroism at \ch{Cu} \textit{K}-edge \uppercase{XANES}},\ }\href
  {https://doi.org/10.1007/s10948-017-4418-5} {\bibfield  {journal} {\bibinfo
  {journal} {Journal of Superconductivity and Novel Magnetism}\ }\textbf
  {\bibinfo {volume} {31}},\ \bibinfo {pages} {663} (\bibinfo {year}
  {2018})}\BibitemShut {NoStop}%
\end{thebibliography}%

\end{document}